\documentclass[aps,twocolumn,superscriptaddress]{revtex4}  
\usepackage[utf8]{inputenc}
\usepackage{graphicx}
\usepackage{comment}

\begin{document}

\title{Tunable intracellular transport on converging microtubule morphologies}

\author{Niranjan Sarpangala}
\affiliation{University of California, Merced, CA, 95343}

\author{Brooke Randell}
\affiliation{California Polytechnic State University, San Luis Obispo, CA, 93407}

\author{Ajay Gopinathan}
\affiliation{University of California, Merced, CA, 95343}

\author{Oleg Kogan}
\affiliation{California Polytechnic State University, San Luis Obispo, CA, 93407}
\email{okogan@calpoly.edu}

\begin{abstract}
A common type of cytoskeletal morphology involves multiple converging microbutubules with their minus ends collected and stabilized by a microtubule organizing center (MTOC) in the interior of the cell.  This arrangement enables the ballistic transport of cargo bound to microtubules, both dynein mediated transport towards the MTOC and kinesin mediated transport away from it, interspersed with diffusion for unbound cargo-motor complexes. Spatial and temporal positioning of the MTOC allows for bidirectional transport towards and away from specific organelles and locations within the cell and also the sequestering and subsequent dispersal of dynein transported cargo. The general principles governing dynamics, efficiency and tunability of such transport in the MTOC vicinity is not fully understood.
To address this, we develop a one-dimensional model that includes advective transport towards an attractor (such as the MTOC), and  diffusive transport that allows particles to reach absorbing boundaries (such as cellular membranes). We calculated the mean first passage time (MFPT) for cargo to reach the boundaries as a measure of the effectiveness of sequestering (large MFPT) and diffusive dispersal (low MFPT). We show that the MFPT experiences a dramatic growth in magnitude, transitioning from a low to high MFPT regime (dispersal to sequestering) over a window of cargo attachment/detachment rates that is close to {\it in vivo} values. Furthermore, we find that increasing either the attachment or detachment rate, while fixing the other, can result in optimal dispersal when the attractor is placed asymmetrically. Finally, we also describe a regime of rare events where the  MFPT scales exponentially with advective velocity towards the attractor and the escape location becomes exponentially sensitive to the attractor positioning. Taken together, our results suggest that structures such as the MTOC allow for the sensitive control of the spatial and temporal features of transport and corresponding function under physiological conditions. 
\end{abstract}

\maketitle

\section*{Introduction}
The transport of material within eukaryotic cells is a critically important physiological process that cannot be achieved by passive diffusion alone. In these cells, cargo, including vesicles and organelles, are dragged along by a variety of molecular motors which utilize energy from ATP hydrolysis to power their directed stepping motion along cytoskeletal protein filaments with a well-defined polarity \cite{howard2002mechanics}.  Motors from different families such as kinesins and myosins step along different filaments (microtubules and actin respectively) and others such as dynein move along the same microtubule filaments as kinesins but in the opposite direction.  Transport at the cellular scale is therefore a complex process that involves phases of multiple motors effecting directed transport along cytoskeletal filament networks interspersed with passive diffusion of the cargo \cite{jenny_review, Koslover_review}. This process is essential for the transport of a variety of cargo between specific locations and organelles within the cell. Examples include the transport of cargo in cilia \cite{Cilia}, between the plasma membrane and Golgi apparatus \cite{Golgi1}, \cite{Golgi2}, between Endoplasmic Reticulum and Golgi \cite{Golgi3}, \cite{Koslover_review}, transport of viruses towards replication sites \cite{Virus1}, \cite{Virus2}, and the transport of many other vesicles and organelles for various functional purposes (see review \cite{Koslover_review}), \cite{Interactome}.

Much like the design of road networks affect traffic flow, the morphologies of the cytoskeletal networks in cells have been shown to have a significant effect on intracellular transport \cite{Ando_Gopinathan, maelfeyt2019anomalous, maelfeyt2019cytoskeletal, hafner2018spatial}. This is particularly important as, even  a single type of cytoskeletal filament such as microtubules exhibit a wide diversity of morphologies within different cell types to enable different functions\cite{MicrotubuleMorphologies}.  In some situations, such as in melanophores microtubules have a strongly orderly (in this case radial) - organization \cite{Melanophores1}.  In others, the orientation or polarity of microtubule (MT) morphology can be broadly distributed.  In pancreatic $\beta$ cells, for example, MTs are arranged with both an orientational and polarity disorder \cite{Beta}, although there is an average polarity.  On the other hand, MTs in neuronal dendrites are essentially aligned with the long direction of the dendrite, but their polarity is not uniform \cite{Dendrite_microtubules} resulting in junctions of plus or minus ends along the dendrite.  
\begin{figure}[h!]
\center \includegraphics[width=3.5in]{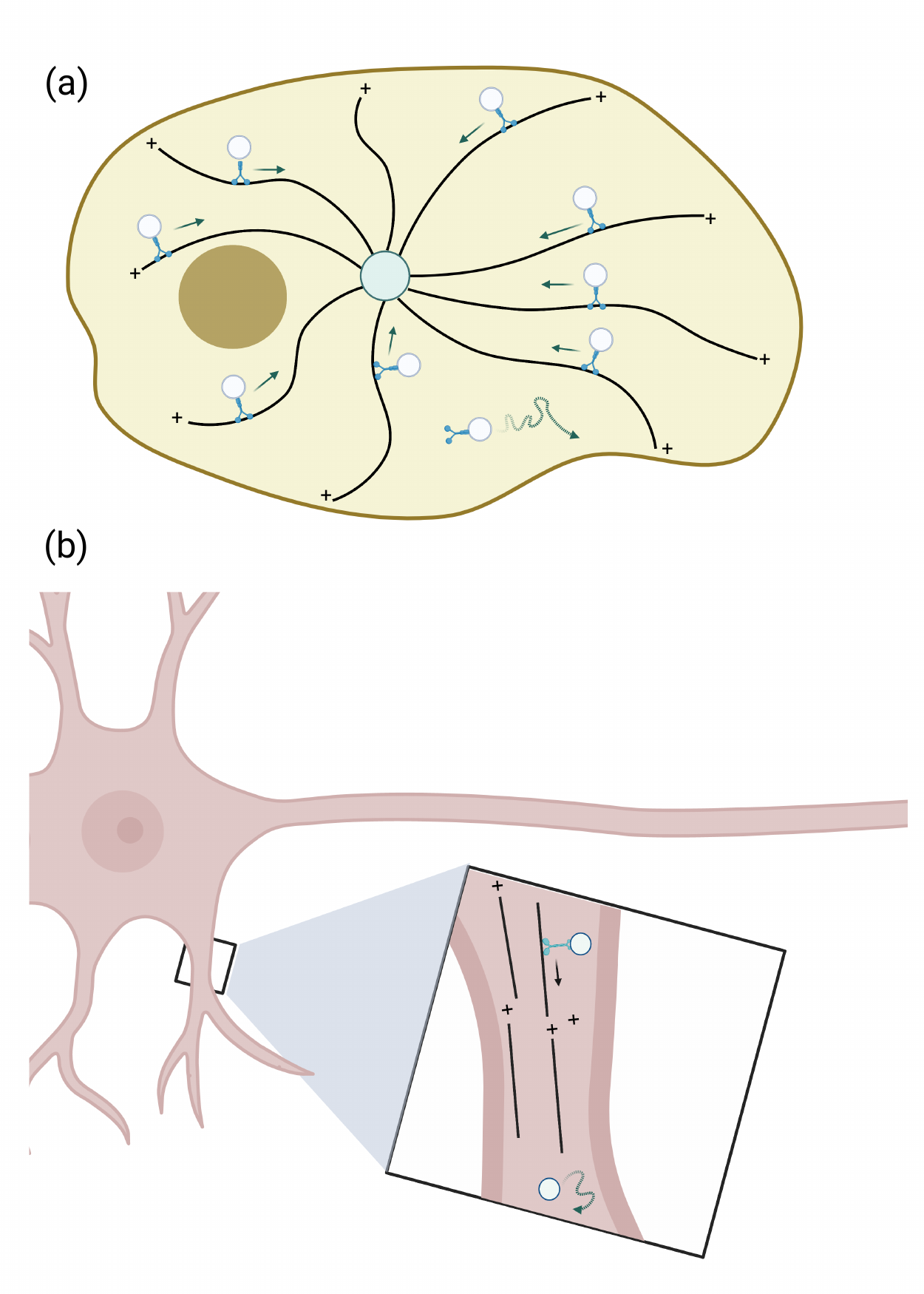}
\caption{(a) A model of a cell in which microtubules have a strong central organization, with minus ends at the centrosome.  A dark circle represents an organelle.  Dynein  motors are shown moving on microtubules.  
(b) One dimensional morphology found in dentrites.  Here the ends of the same polarity from different microtubules can face each other.  This schematic is based on \cite{masucci2022microtubule}. }
\label{fig:cartoon_Niranjan}
\end{figure}

A common structural feature that governs these microtubule morphologies is the microtubule organizing center (MTOC) that is responsible for growing MTs and localizing and stabilizing their minus ends leading to multiple MTs converging with their minus ends at the MTOC \cite{MicrotubuleMorphologies}.  Dynein-driven transport along MTs will move cargo to the vicinity of MTOC, while kinesin mediated transport moves cargo away from it. These ballistic phases are interspersed with isotropic diffusion for unbound cargo-motor complexes. The spatial and temporal positioning of the MTOC therefore allows for bidirectional transport towards and away from specific organelles that can act as MTOCs as well as locations within the cell in the vicinity of the MTOC. Examples in which MTOC facilitates direct transport to the destination of interest include transport of cargo such as secretory vesicles away from the Golgi apparatus toward the cell membrane and endocytic vesicles towards the Golgi which is known to perform as an MTOC in many mammalian cells \cite{Golgi1}, \cite{Golgi2}. The dynein mediated transport of some viruses toward the nuclear envelope is also enabled by the presence of  
 a MTOC  in the vicinity of the nucleus \cite{Virus1, 
 Virus2}. 
 
In some cases, cargo need to traverse regions with convergent MT 
morphologies. Such cases occur in dendritic processes of neuronal cells that have been shown to have regions of alternating polarity of MTs \cite{Dendrite_microtubules}. Directed transport of dynein (kinesin) carrying cargo at a junction of minus(plus) ends will have to overcome what is essentially a trap to maintain observed unidirectional transport towards or away from the main cell body \cite{Dendrite_microtubules}. 

Finally, the location of MTOCs can also be tuned over time to accommodate different cellular functions such as sequestering and dispersal of cargo.  For example, in melanophores \cite{Melanophores1, Melanophores2}, a perinuclear MTOC produces a radial MT structure with minus ends in towards the nucleus and plus ends out toward the membrane. Cells achieve color change by aggregating and sequestering  pigment containing melanosomes near the nucleus via ballistic dynein mediated transport. Upon hormonal stimulation they can switch to a superdiffusive dispersal phase powered by a combination of kinesin and actin. Another example occurs in lymphocytes that enable cytotoxicity by secreting the contents of lysosomes (lytic granules) at the immunological synapse to kill the target cell. Here, dynein dependent sequestering of the lytic granules at the MTOC occurs rapidly followed by the gradual movement of the MTOC towards the synapse with subsequent secretion \cite{nath2016dynein, mentlik2010rapid}.   

In all these cases, it is important to understand the dynamics of the transport and its sensitivity to biological parameters in order to understand functional efficiency and robustness. In particular, given the wide variety of functional contexts in which the converging MT geometry facilitates transport, it is critical to understand the general principles governing dynamics, efficiency and tunability of such transport in the MTOC vicinity.
\\
 
To address this gap, we develop a simple one-dimensional model that includes advective transport towards an attractor (such as the MTOC), and  diffusive transport that allows particles to reach absorbing boundaries (such as cellular membranes). This can be viewed as a 2-layer model consisting of an advective layer endowed with an attractor, a diffusive layer, and absorbing boundaries along the perimeter of the domain. We take the mean first passage time (MFPT) for cargo to reach the boundaries as a measure of the effectiveness of sequestering or directed transport (large MFPT) and diffusive dispersal (low MFPT).
 The number of independent control parameters in this problem can be reduced to four.  These are the rates of attachment to and detachment from microtubules, advective velocity, and the placement of the attractor within the domain.

Using this model we were able to make a series of tantalizing predictions - on which we report here.  A central calculation here is the residence time, or what is commonly called in the literature the mean first passage time (MFPT).  Thus, given an initial location of the cargo within the domain (determined by organelle placement), this quantity tells the average time to reach either of the absorbing boundaries (i.e. escape the domain), or a specific boundary (in one dimension, left or right).  Another relevant quantity is the probability of escape through one or the other domain.  

Symmetric, or nearly symmetric attractor positions can give rise to a dramatic increase in the value of MFPT within a certain window of dimensionless coupling rates between the layers.  Concurrently with this dramatic rise of MFPT, the probability to escape purely diffusively goes to zero in the same range of (dimensionless) coupling values.  This means that for larger coupling values, any cargo particle will have to experience at least one episode of motion on microtubules.  Crucially, we found that biophysical parameters in cells correspond precisely to this range of dimensionless coupling rates.  This suggests that parameter values in cells are optimized for the greatest sensitivity to small changes.  With such parameters, a cell can achieve the largest change in functionality with smallest changes in parameter values.

Second, we predict the existence of optimal coupling rates that minimize the MFPT.  This minimal MFPT happens when the attractor is positioned asymmetrically (off center) in the domain.  A similar phenomenon has been predicted in the study of diffusion with stochastic reset \cite{Reset1}, \cite{Reset2}.  Indeed, attachment to the microtubule, followed by a rapid transport to the attractor, followed by detachment from the microtubule back to diffusion in the cytoplasm is effectively a reset.  

When the coupling rates are much larger than all other rates in the problem, the model reduces to effectively one-layer.  Here we demonstrate that even a slight asymmetry in the position of the attractor can lead to a very strong amplification of the preferred exit end.  
This provides another example of sensitivity to small parameter changes - in this case asymmetric of the attractor placement.  This effect happens at sufficiently large advective velocity, and corresponds to rare event physics.  In the regime of rare events, a small fraction of particles escape quickly, while the majority advect to the attractor, and form a quasi-stationary distribution around it.  They stay in the vicinity of the attractor for a time that scales exponentially with advective velocity (or inverse of diffusion coefficient).  


\section*{Methods}
\subsection*{Model}
We consider the minimal  model in a one-dimensional domain of length $L$.  It contains an advective layer (AL) that represents motion along microtubules, and a diffusive layer (DL) that represents diffusion in the cytoplasm.  We assume that attachment to and detachment from microtubules are Poisson processes, endowed with rates $\alpha$ and $\beta$ respectively.  This means, for example, that a motor spends on average a time $1/\beta$ since attaching to a microtubule.  While advecting, particles move with a uniform velocity towards the attractor - which is an attracting fixed point located at some coordinate $x=X_0$ between $x=0$ and $x=L$.  Letting $\rho(x)$ and $\theta(x)$ be probability densities of particles in the AL and DL respectively, the model reads
\begin{eqnarray}
\frac{\partial \rho}{\partial t} &=& -\frac{\partial}{\partial x}\left(v(x)\rho\right) + \alpha \theta - \beta\rho \\
\frac{\partial \theta}{\partial t} &=& -\alpha\theta + \beta\rho + D \frac{\partial^2 \theta}{\partial x^2}
\end{eqnarray}
on $0\leq x \leq L$.  The velocity field is given by 
\begin{equation}
v(x) = \left\{\begin{array}{c} +v_0 \mbox{ ... } x<X_0 \\ -v_0 \mbox{ ... } x>X_0 \end{array} \right.
\end{equation}
The parameters are rates $\alpha$ and $\beta$, the diffusion coefficient $D$, the advective velocity on microtubules $v_0$, and the location of the attractor $X_0$.  There are absorbing BCs at $x=0$  and $x=L$, i.e. $\rho(0) = \theta(0) = 0$ and $\rho(L)=\theta(L)=0$.  All together, there are six physical parameters.

\begin{figure}[h!]
\center \includegraphics[width=\linewidth]{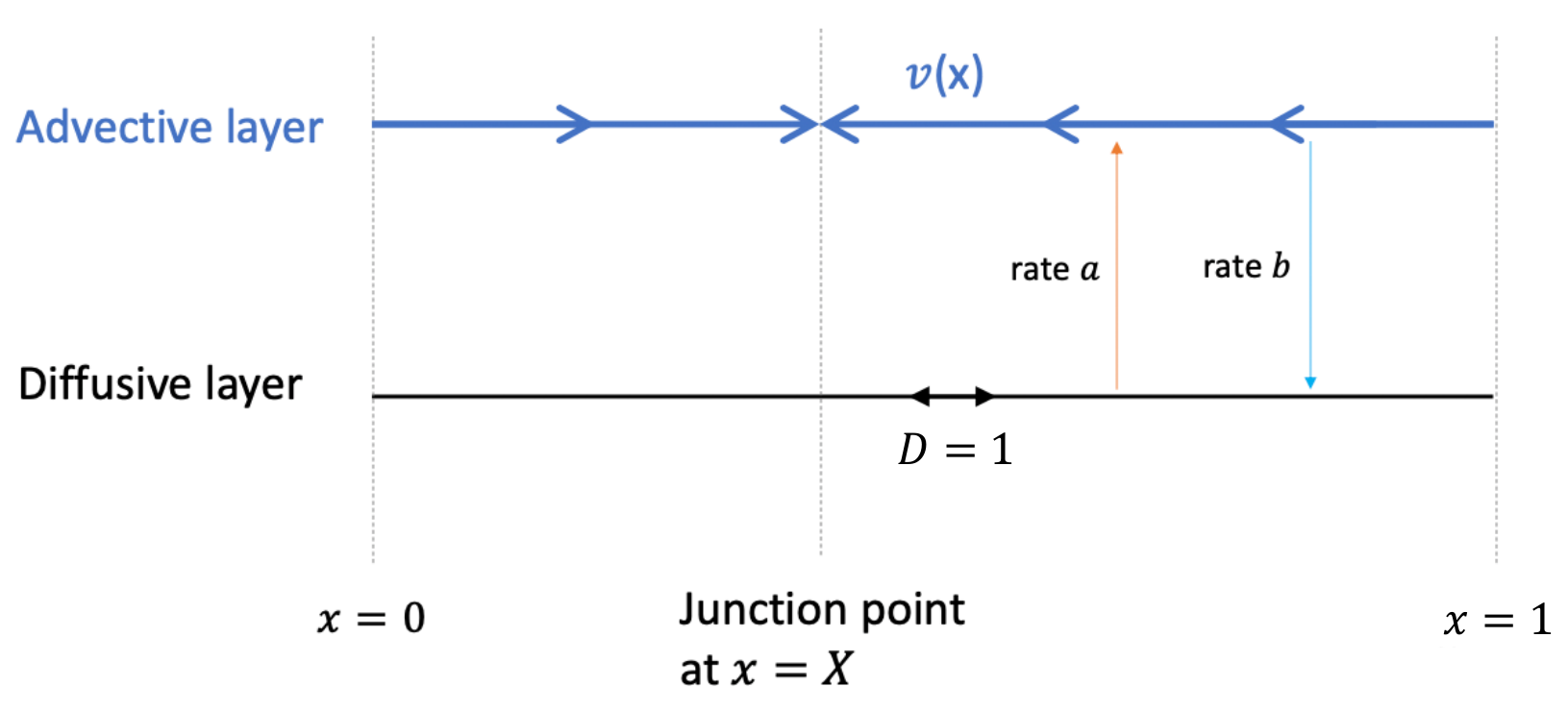}
\caption{One-dimensional model with dimensionless parameters.}
\label{fig:Model_full}
\end{figure}

We will switch to dimensionless variables by rescaling the lengths by $L$ and times by $L^2/D$. Thus, $x' = x/L$ and $t' = tD/L^2$.  The resulting equations will be 
\begin{eqnarray}
\label{eq:RhoEq} \frac{\partial \rho}{\partial t'} &=& -\frac{\partial}{\partial x'}\left(v'(x)\rho\right) + a \theta - b\rho \\
\label{eq:ThetaEq}\frac{\partial \theta}{\partial t'} &=& -a\theta + b\rho +  \frac{\partial^2 \theta}{\partial x'^2}
\end{eqnarray}
on $0<x'<1$, with $\rho(0) = \theta(0) = 0$ and $\rho(1) = \theta(1) = 0$, the velocity field 
\begin{displaymath}
v'(x) = \left\{\begin{array}{c} +v \mbox{ ... } x<X \\ -v \mbox{ ... } x>X \end{array} \right.
\end{displaymath}
where $X = X_0/L$ and $v = \frac{v_0 L}{D}$, and coupling rates $a = \frac{\alpha L^2}{D}$ and $b = \frac{\beta L^2}{D}$.  From now on, we will drop primes.  The model is depicted schematically in Fig.~\ref{fig:Model_full}.
\subsubsection*{Range of parameters}
\label{sec:Estimations}
Here we review the values of parameters from literature.  Both adsorption rate $\alpha$ and desorption rate $\beta$ are expected to be of the order of $1$ per second.  For example, \cite{Ando_Gopinathan} cites $\alpha = 5$ $s^{-1}$ and $\beta = 1$ $s^{-1}$.  Microtubule lengths typically fall in the range of $1-10$ $\mu m$ \cite{Ando_Gopinathan}.  However, the length of advective path may be much larger.  For example, in neurons, a cargo that needs to be delivered from the soma to synapses on the ends of axons will travel a length of the order of a meter \cite{Koslover_review}.  The velocity of molecular motors on MTs is on the order of $1$ $\mu m/s$ \cite{Koslover_review}, \cite{Ando_Gopinathan}, although this quantity also has a degree of variability \cite{Velocity_tuning}.  Diffusion coefficient of vesicular organelles in the cytoplasm fall in the range $10^{-3} - 10^{-1}$ $\mu m^2/s$ \cite{Koslover_review}.

Given these physical parameters, our dimensionless parameters $a$ and $b$ will take on values in the range $[10, 10^5]$, and parameter $v$ will take on values in the range $[10,10^4]$. 

There are four timescales in the problem: $1/a$, $1/b$, the advective timescale $1/v$, and the diffusive timescale (which is of order $1$ in dimensionless units).  Different special cases or behavioral regimes emerge when one of these timescales differs significantly from others.  

The limit that is particularly amenable to analysis is one in which $1/a$ and $1/b$ are both much smaller than the advective time (which is of order $1/v$ in dimensionless units) and diffusive time (which is of  order $1$ in dimensionless units).  We will formally call it the $a,b \rightarrow \infty$ limit.  In this regime,  the model reduces to an advection-diffusion process in one single layer, which amenable to many analytical results.  
 
\subsection*{Analytical approach in the one-layer limit}
A very important special case is $a=b$.  As $a=b \rightarrow \infty$, the model reduces to an effeective one-layer model:
\begin{displaymath}
\frac{\partial P}{\partial t} = -\frac{\partial}{\partial x}\left(v(x)P(x) - \frac{\partial P}{\partial x}\right)
\end{displaymath}
where $P(x,t)$ is the probability density (i.e. $P$ describes both $\theta$ and $\rho$, which become identical).  A general solution will be written as an eigenfunction expansion
\begin{equation}
\label{eq:series_soln_appendix}
P(x,t) = \sum_{n} c_n p_n(x)e^{\sigma_n t},
\end{equation}
where $p_n(x)$ and $\sigma_n$ is $n$th eigenfunction and eigenvalue, which satisfy $Op_n = \sigma_n p_n$, with the operator $O$ given by 
\begin{equation}
O = -\frac{\partial}{\partial x}\left(v(x) - \frac{\partial }{\partial x}\right),
\end{equation}
with
\begin{equation}
\label{eq:adv_vel}
v(x) = \left\{\begin{array}{c} +v \mbox{ ... } x<X \\ -v \mbox{ ... } x>X \end{array} \right.
\end{equation}
and a constant $v$.  Thus, the one-layer model contains two parameters: dimensionless advective velocity $v$ and dimensionless position of the attractor $X$, which can take on values between $0$ and $1$.

The computation of eigenvalues $\sigma_n$ and eigenfunctions $p_n(x)$ of the operator $O$, as well as the computation of the  eigenfunctions $q_n(x)$ of the operator $O^{\dag}$ is given in Appendix B.

Starting from the initial condition $P(x,t=0) = \delta(x-x_0)$, the probability density will be given by 
\begin{equation}
P(x,t; x_0) = \sum_{n} \frac{q^*_n(x_0)p_n(x)}{\int_0^1 q^*_n(x')p_n(x')\,dx'} e^{\sigma_n t}
\label{eq:P-sum}
\end{equation}
Everything that we need to compute MFPT can be extracted from this probability density.

To calculate the MFPT $\tau(x_0)$, we notice that the magnitude of the flux through the boundary is given by $f(t) = \left| \frac{\partial P}{\partial x}\right |_{bdry}$ in dimensionless units.  Then $f(t)dt$ gives the fraction of initial particles that cross the boundary in $[t,t+dt]$ = probability of crossing that boundary in $[t,t+dt]$, since the initial condition is normalized to $1$.  So, $p=\int_0^{\infty} f(t)\,dt$ gives the probability of ever leaving through that boundary, $\frac{f(t)dt}{p}$ gives the probability that particles that leave through that boundary do so in $[t,t+dt]$, and finally $\tau  = \int_0^{\infty} t\frac{f(t)}{p} \,dt$ is the average time to leave through that boundary.  In this problem, there are two boundaries, with $\tau_l$ and $\tau_r$ denoting MFPT to exit through the left and the right boundary respectively.  We expect $\tau_l \rightarrow 0$ as $x_0 \rightarrow 0$ and $\tau_r \rightarrow 0$ as $x_0 \rightarrow 1$.  Finally, MFPT in general - without conditioning on a specific boundary - is the weighted average of the two: $\tau  = \tau_l p_l + \tau_r p_r$, which matches predictions using other methods \cite{gardiner1985handbook}.

\subsection*{Analytical approach in the general case}
\label{sec:theory_general}
Analogously to the one-layer approach, we again seek a general solution to Eqs.~(\ref{eq:RhoEq})-(\ref{eq:ThetaEq}) via an eigenfunction expansion of the form
\begin{equation}
\label{eq:expansion_general}
\left(\begin{array}{c} \rho(x,t) \\ \theta(x,t) \end{array}\right) = \sum_n c_n  \left(\begin{array}{c} R_n(x) \\ \Theta_n(x) \end{array}\right)e^{-\sigma_n t}
\end{equation}
(we found it convenient to factor out the negative sign from $\sigma$ here), where $\left(\begin{array}{c} R_n \\ \Theta_n \end{array}\right)$ and $\sigma_n$ is the $n$th (vector) eigenfunction and eigenvalue, which satisfy $O\left(\begin{array}{c} R_n \\ \Theta_n \end{array}\right) = \sigma_n\left(\begin{array}{c} R_n \\ \Theta_n \end{array}\right)$, with the operator $O$ given by 
\begin{equation}
O = \left(\begin{array}{cc} \frac{\partial}{\partial x} v(x) + b & -a \\ -b & a - \frac{\partial^2}{\partial x^2} \end{array}\right)
\end{equation}
with $v(x)$ given by Eq.~(\ref{eq:adv_vel}).  The full model contains four parameters: dimensionless advective velocity $v$, dimensionless rates $a$ and $b$, and dimensionless position of the attractor $X$, which can take on values between $0$ and $1$.  The computation of eigenvalues and eigenfunctions is given in Appendix A.

Remarkably, there are only a finite number of eigenfunctions and eigenvalues.  In other words, the eigenset is not complete. As $a=b\rightarrow \infty$, this number goes to infinity, while the lower-lying eigenvalues and eigenfunctions approach those of the one-layer model.  The completeness is not guaranteed, since the operator $O$ is not Hermitian.  Thus, an expansion such as in Eq.~(\ref{eq:expansion_general}) is of limited use, and cannot be used to fit a solution for an arbitrary initial condition - including a point-like $\delta$ function initial condition.  This also implies that we cannot compute escape currents and MFPT from such initial conditions.

However, we can always compute the ground state eigenvalue, $\sigma_1$.  Then the time $1/\sigma_1$, while not a true MFPT, is an estimate of a characteristic time for escape.  We found  that this time alone agrees with MFPT computed in simulations quite well, so we will make MFPT arguments based on this estimate.
\subsection*{Monte Carlo simulation method}
\label{sec:Simulations}
We considered a simple one dimensional computational model to simulate the transport process in a domain of length L with attractor formed by oppositely oriented microtubules. Our computational model involves two layers, an advective layer (AL) where the particle undergoes active transport and a diffusive layer where it does one dimensional random walk. We consider one particle at a time. To begin, we initialize the particle at position $x=x_0$ within the domain $x \in [0, L=1]$ either in the diffusive or in advective layer as required.  We consider that the particle can switch from diffusive layer to advective layer with a rate $a$ and from advective layer to diffusive layer with a rate $b$. When a particle switches to diffusive layer, a time $t_d$ is drawn from the exponential distribution $e^{-at}$ and the particle is allowed to diffuse for $n=t_d/\Delta t$ number of steps. $\Delta t$ is the time step in the simulation. In each step the position is updated as
\begin{equation}
x(t+\Delta t)=x(t)+r \Delta x,
\end{equation}
where $r$ is drawn from the set $\{-1, 0, 1\}$ with the probability $p=1/ 3$. $\Delta x$ is the step size which is chosen such that the diffusion constant of the particle $D=\frac{p\Delta x^2}{\Delta t}$ is  $1$.  Right after finishing a diffusive portion of a simulation run, the particle switches from diffusive to advective layer. In the advective layer, the particle stays for a time $t_a$ drawn from  $e^{-bt}$, \textit{i.e.} $n= t_a/\Delta t$ number of steps. In the advective layer, the position of the particle is updated as 
\begin{equation}
x(t+\Delta t)=x(t)+v(x)\Delta t,
\end{equation}
where $v(x)$ is the advective velocity given by Eq.~(\ref{eq:adv_vel}).
These alternative portions of a simulation run in diffusive and advective layers are continued until the particle reaches one of the boundaries ($x=0 $ or $x=1$) or until maximum simulation time, $T_{max}$ is reached.  We then repeat with $N$ particles to get enough statistics to calculate the overall MFPT, probabilities and MFPTs to exit out of specific boundaries, and other quantities.
\subsubsection*{Trajectories}
\begin{figure}[h!]
\center \includegraphics[width=\linewidth]{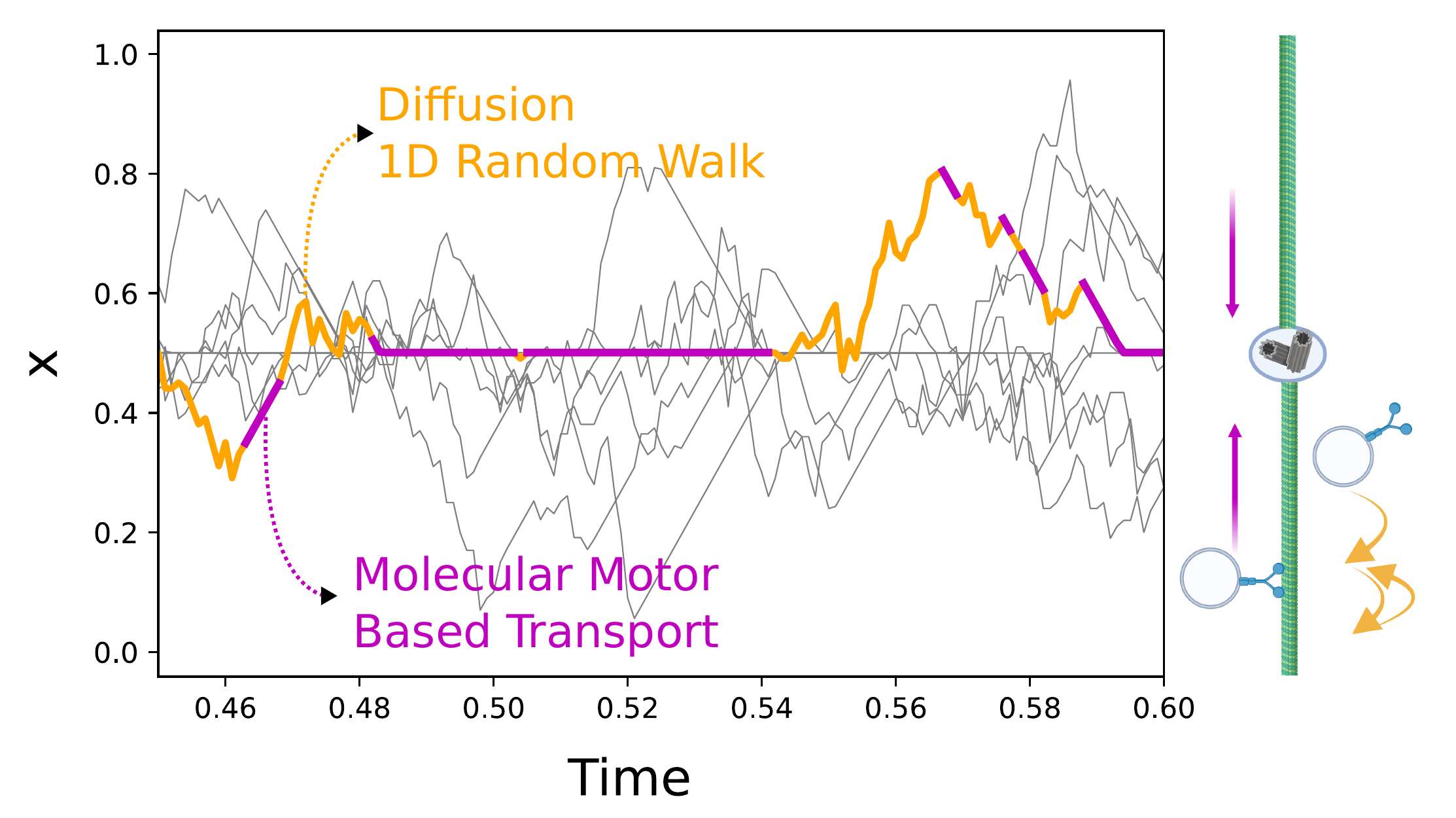}
\caption{A sample trajectory generated by the Monte Carlo simulation. Diffusive motion is indicated with orange line, and advective motion with a magenta line. Grey colored lines indicate more sample trajectories. Here $X=0.5$. }
\label{fig:one_trajectory_sample}
\end{figure}

To get the trajectories, we record the data of the $x$ position and the layer in which particle is located at regular time intervals during each simulation run.  An example of trajectories is shown in Fig.~\ref{fig:one_trajectory_sample}.

\subsubsection*{Computation of Net MFPT}
To compute the net MFPT for a given parameter set, we perform simulation runs until the particle exits out of one of the boundaries ($x=0$ or $x=1$). We record the time of exit for each run and then compute the mean and standard error of the mean for all $N$ runs.

\subsubsection*{Computation of Conditional MFPT and escape probability}
To compute the MFPT for exit specifically through the left (or the right) boundary, we record the time as well as the boundary through which the particle exits. Then we filter out only those simulation runs where a particle exited out of the left (or right) boundary. Then we compute mean and standard error of the mean for those runs. We compute the escape probability through left (or right) boundary as the fraction of runs that exited out of the left (or right) boundary.

\subsubsection*{Statistics of visits to the AL}
We measure the fraction of simulation runs in which a particle that started on the DL ended up making at least one visit to the AL.  In each simulation run, we also compute the number of visits to the advection layer before exiting. To do this, we update a counter every time the particle switches its layer to get the number of times it switches layers prior to exiting the domain.  We then compute the average over $N$ runs.

\section*{Results and Discussion}
\subsection*{Variation of coupling rates can change escape times by orders of magnitude}
We begin our presentation of results with the symmetric case, $X=1/2$. 
\begin{figure}[h!]
\center \includegraphics[width=3.6in]{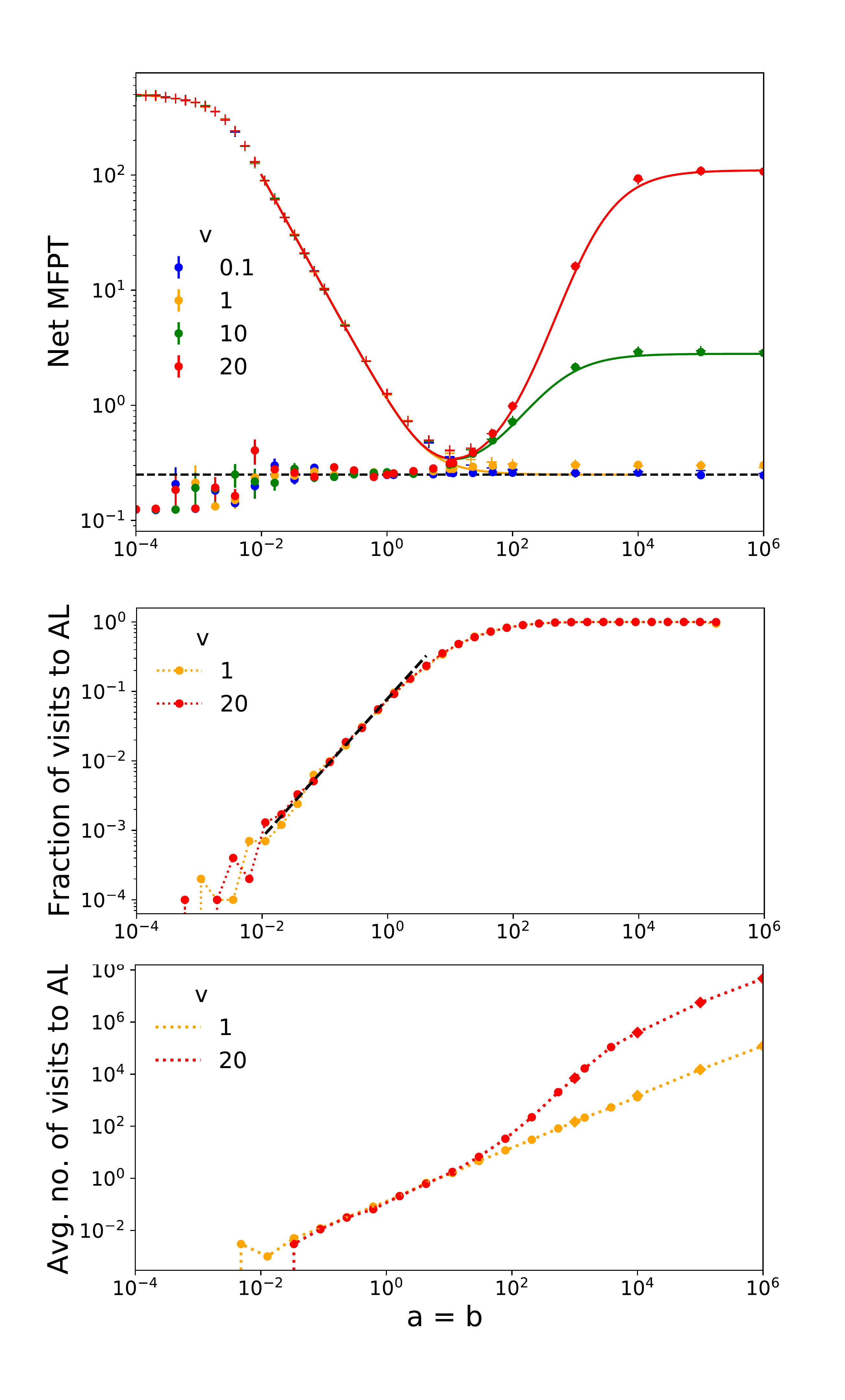}
\caption{Symmetric case: $X=0.5$, the initial location of particles is also at $x_0 = 0.5$. (a) MFPT vs.~$a=b$.  Dots: IC on the DL; crosses: IC on the AL.   The solid curves are analytical estimations of MFPT given by $1/\sigma_1$, where $\sigma_1$ is the ground state eigenvalue.  The MFPT is in dimensionless time units; to convert to time in seconds, multiply by $L^2/D$ expressed in physical units.  The dashed horizontal line has a value $0.25$. The last two points ($a=10^5$ and $10^6$) required a smaller $dt=\frac{10^{-6}}{3}$; $dt = \frac{10^{-4}}{3}$ was sufficient for the rest.  Therefore, we used $N=10^3$ for the last two points to optimize simulation time, and $N=10^4$ for the rest.  (b) Fraction of simulation runs that visit the AL at least once after starting in the DL.  The dashed line is a fit, of the form $0.079 a$.  Here $N=10^4$ and $dt = \frac{10^{-4}}{3}$. (c) Average number of visits for particles starting in the DL.  Here $N=10^3$, $dt=\frac{10^{-4}}{3}$ (circles), and $\frac{10^{-6}}{3}$ (diamonds).  The $x$-axis is the same in all three plots; the plots are aligned.  The shading guides the eye to the second crossover region.}
\label{fig:NiranjunSym}
\end{figure}
For simplicity we will set the particles' initial placement at $x_0=1/2$ - this is the initial condition (IC) in analytical calculations - and let $a=b$ for now.  Figure \ref{fig:NiranjunSym} (a) displays the mean first passage time (MFPT) as a function of $a=b$ at different advective speeds $v$.  To help understand the physics of the process, we also plot the fraction of times that particles visit the advective layer  in panel (b) (for particles initially placed on the DL), as well as the number of times they do so in panel (c) (also when starting on DL).  

Two crossovers are evident from the plot of MFPT vs $a \mbox{ }(=b)$.  The first crossover takes place around $a=10^{-2}$.  As suggested by the plot of the fraction of visits to the AL, at this coupling rate the probability of visiting the AL becomes non-zero; below this crossover, the advective layer is not visited and the MFPT is a purely diffusive time $\approx 0.12$.  For $a$ above this crossover value,  the probability of visiting the AL grows with increasing $a$.   While the fraction of particles visiting the AL grows  $\propto a$, the time to remain in the AL (the longest time scale in this range of $a$) decreases $\propto 1/a$, resulting in the plateau of MFPT vs.~$a$.  Because the probability (or fraction) to visits to the AL is less than $1$ (for particles startin in the DL), a particle has a chance to escape purely diffusively for $a$s in this plateau region.  

The MFPT is in dimensionless time units; to convert to time in seconds, multiply by $L^2/D$ expressed in physical units.  For example, for $L=1$ $\mu m$ and $D= 10^{-2}$ $\mu m^2/s$, the MFPT of $10$ dimensionless time units corresponds to $10^3$ seconds.  The MFPT for diffusive transport on a domain with two absorbing boundaries and a midpoint initial condition is $0.125$ (in dimensionless time units), which is half of the first plateau value, and much lower than plateaus after the second crossover for $v>1$.

We continue our discussion of Fig.~\ref{fig:NiranjunSym}. The probability of visiting the AL (for particles starting in the DL) eventually reaches $1$ at larger $a$;  particles are now certain to visit the AL at least once.  In other words, the probability of a purely diffusive escape reaches zero and we encounter the second crossover.  For $v=20$, for example, this second crossover happens around $a=b=10$, but its location - defined by the point of inflection - varies somewhat with $v$.  This crossover is broad - it can be several decades wide - and marked by a drastic growth of the MFPT, especially at larger $v$.  In this second crossover regime, each particle experiences intermittent advection, punctuated by periods of diffusion.  In other words, on a typical run from an initial location to one of the boundaries, a particle's trajectory will include multiple episodes of advection and diffusion following each other.  Eventually, we reach the second plateau, when the switching between the layers is so rapid that we now reach an effectively one-layer regime.  This regime will be studied in the next section, where we examie a one layer model with advection and diffusion taking place simultaneously.  MFPTs predicted by that model match the high $a=b$ plateaus.  Interestingly, there is a strong velocity dependence in the one-layer regime, but not in the range of $a=b$ in the plateau below the second crossover.  

For $a=b<$ $1/($simulation time$)$, particles with IC in the advective layer (plus symbols in Fig.~\ref{fig:NiranjunSym}) will never enter the DL and therefore will not escape.  MFPT will simply be limited by the simulation time - this is manifested in the saturation at MFPT =  $500$, since this was the simulation time.

Appendix C displays examples of particle trajectories for a broad range of $a=b$ that cover all of the behavioral regimes shown in Fig.~\ref{fig:NiranjunSym}.  These figures  demonstrate the change in the character of trajectories - from the types that contain advective periods long enough to arrive to the attractor at low $a=b$, to intermittent behavior in the second crossover region, to very rapid switching between layers for $a=b$ beyond the second crossover - when the model is effectively in the one-layer regime.

\subsubsection*{The region of the most sensitive behavioral tuning matches the biological parameters}
We now turn our attention to the biological significance of these results.  Note that the second crossover takes place between $a=10$ and $a=10^4$.  Remarkably, this is precisely the range of these parameters found in cells - see ``Range of parameters'' above.  This might imply that these parameters evolved to have such values for an easy tunability.  Indeed, the second crossover region is precisely where a change in the rates gives rise to the largest change in the outcome - especially at larger values of $v$.

\subsection*{There is an optimal coupling rate between advective and diffusive behavior}
\begin{figure}[h!]
\center \includegraphics[width=3.1in]{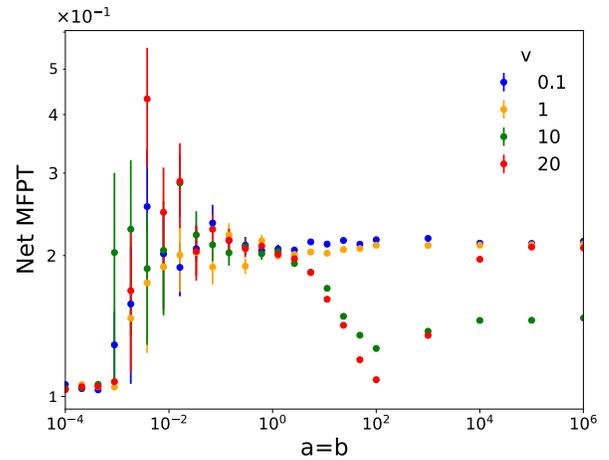}
\caption{Asymmetric case: $X=0.85$.  In this particular case, $x_0=0.7$, but such dips are also present at other $x_0$.}
\label{fig:NiranjunAsym}
\end{figure}

Placing the attractor asymmetrically can give rise to a decrease in MFPT with increasing coupling rates - see Fig.~\ref{fig:NiranjunAsym}.  
This effect is only seen at larger $v$.  The decrease happens over a range of $1/a$ that is comparable to the advective time, $\sim 1/v$.  For example, for $v=20$, the time scale to travel advectively to the attractor is $\sim 0.05$, while the decrease is seen for $a$ between $1$ and $100$,  which corresponds to the time scale between $1$ and $0.01$.  

We think that this decrease in the MFPT happens because an increase in the interlayer coupling causes more material to congregate at the attractor, which is close to one of the ends - thus leading to an overall decrease in the MFPT. 

\begin{figure}[h!]
\center \includegraphics[width=3.4in]{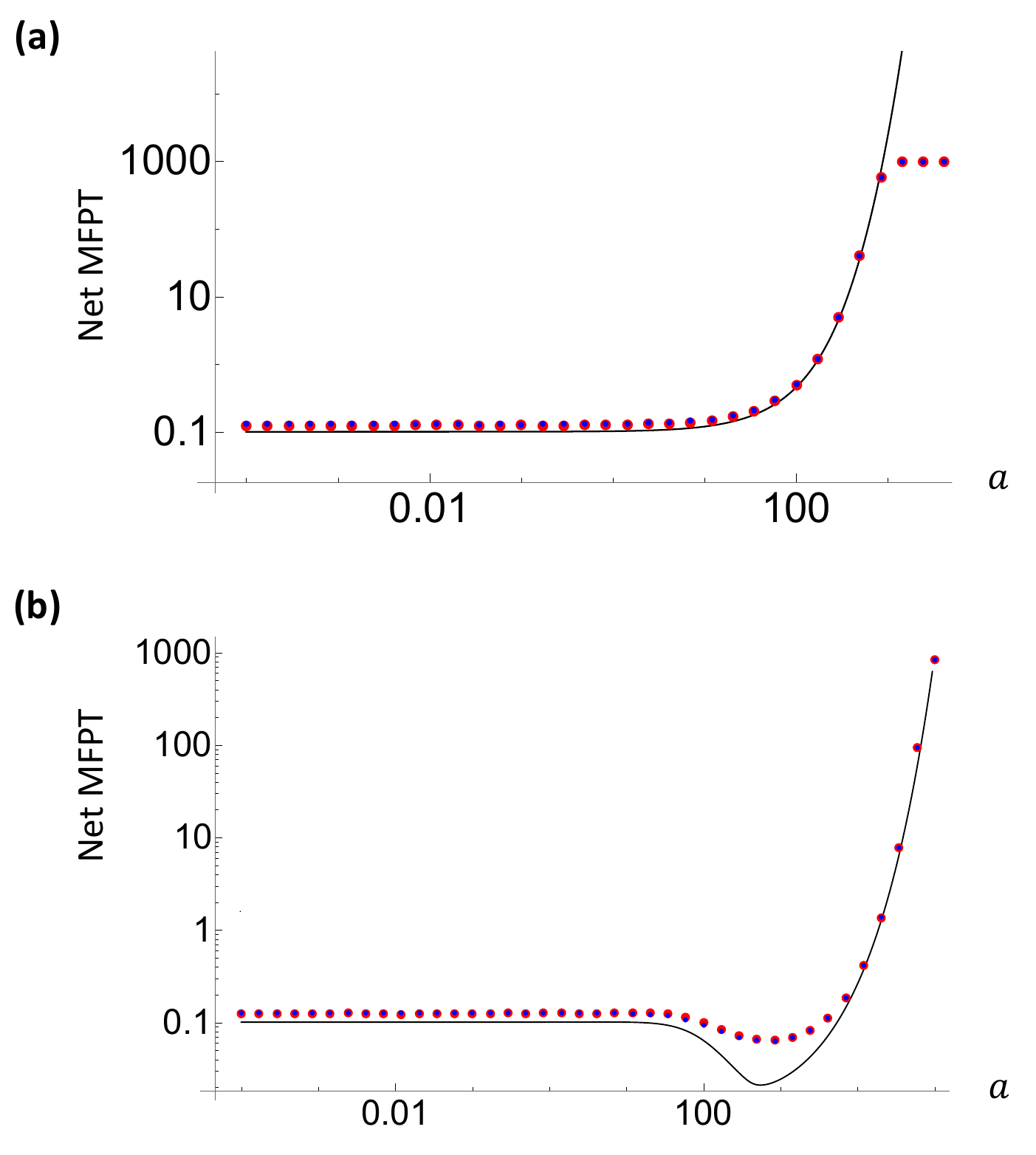}
\caption{$\tau(a)$ at fixed $b=169$.  (a) $X=0.5$, (b) $X=1/26$. $v=13$ for both.   Lines: theory, dots: simulation.  Red dots - IC on the diffusive layer, blue dots - IC on the advective layer.  The numbers for the two types of initial conditions are not identical, but the difference is almost invisible.  The analytical prediction is $1/\sigma_1$ - the inverse of the ground state eigenvalue, which is not a true MFPT.  The IC in the simulation was at $x_0=0.5$.  The simulation time was $1000$, which is the reason for flattening of the simulation data at large $a$ in panel (a).}
\label{fig:DipComparison_vary_a}
\end{figure}

Fig.~\ref{fig:DipComparison_vary_a} shows an example of this phenomenon due to only the parameter $a$ varied at fixed $b$.   
We mentioned in the discussion of the analytical approach in the general case that a complete eigenset in the two-layer model does not exist, so the exact solution cannot be obtained as a sum of the modes.  However, the MFPT can be estimated as $\tau=1/\sigma_1$, where $\sigma_1$ is the ground state eigenvalue.
\begin{figure}[h]
\center \includegraphics[width=3.4in]{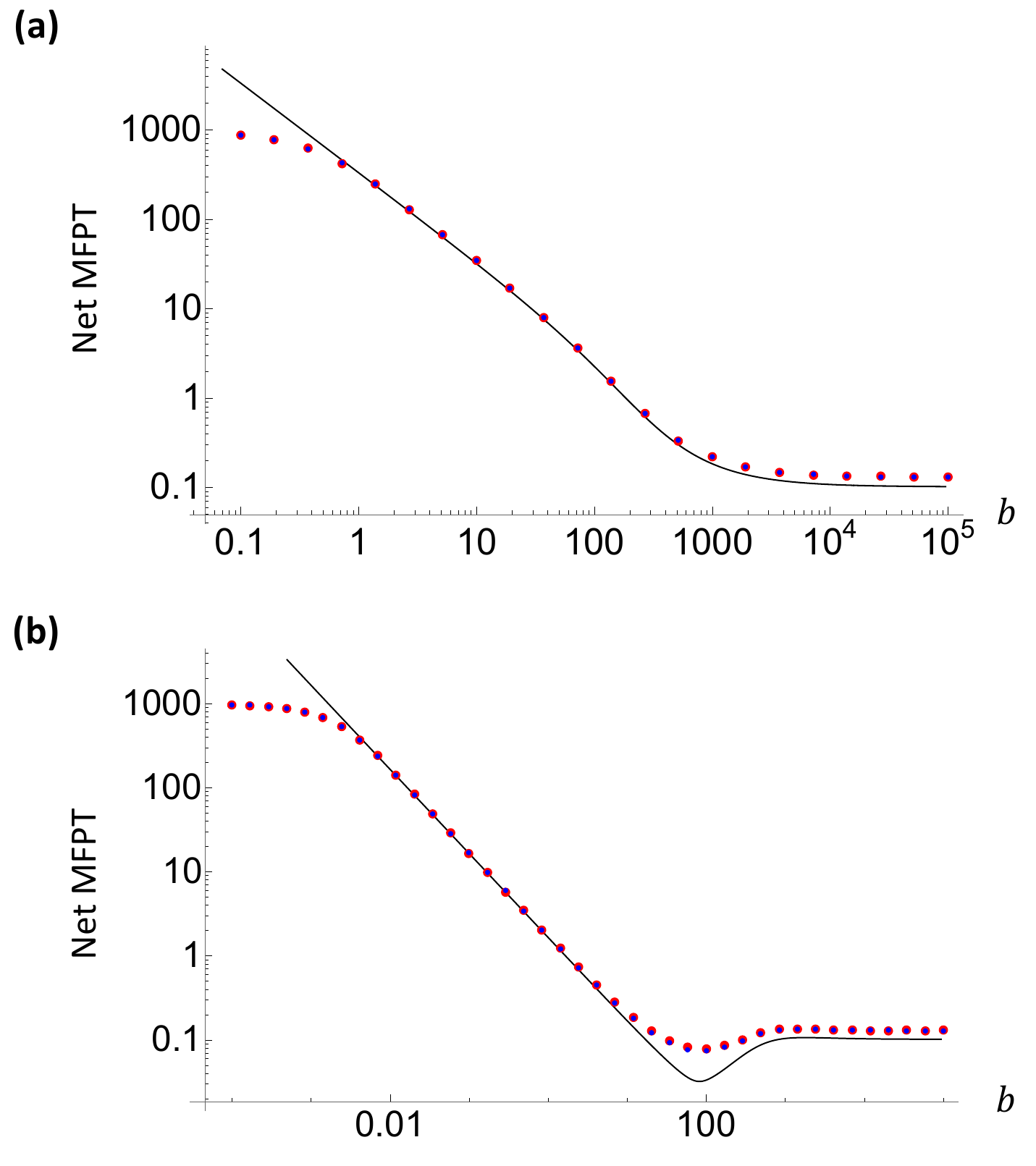}
\caption{$\tau(b)$ at fixed $a=169$. (a) $X=0.5$, (b) $X=1/26$.  $v=13$ for both. 
Lines: theory, dots: simulation.  Red dots - IC on the diffusive layer, blue dots - IC on the advective layer.  The analytical prediction is $1/\sigma_1$ - the inverse of the ground state eigenvalue, which is not a true MFPT.  The IC in the simulation was at $x_0=0.5$.  We again see saturation of simulation results at low $b$ at the simulation time (here, $1000$ time units). 
 }
\label{fig:DipComparison_vary_b}
\end{figure}
The solid lines in Fig.~\ref{fig:DipComparison_vary_a} are values of $1/\sigma_1$.  This estimation should become more accurate as escape events become rare (MFPT $\gg$ than all other time scales); this is because higher eigenmodes contribute little to the probability current in the rare event limit.   Moreover, while this calculation does not give IC dependence, MFPT loses this dependence as escape events become rare.  Some discussion of rare events can be found in the next section, and a much more in-depth discussion will appear in \cite{Our_PRE}.

The dips in Fig.~\ref{fig:DipComparison_vary_a}  happen, again, because increasing $a$ causes particles to return back to the attractor, thus minimizing the chance for them to wander too far to the right while diffusively exploring the long part of the domain.  On the other hand, increasing $a$ even further tends to keep the particles in the AL and therefore prevents them from escaping (particles cannot move in the direction of the ends when they are in the AL due to the advective flow being directed towards the attractor).  

These dips are somewhat counter-intuitive - an overall escape time is lowered by increasing the tendency to go towards the attractor inside the domain - as long as the attractor is placed asymmetrically.   

A similar phenomenon has been reported in connection to the problem of mean first passage time with a reset \cite{Reset1}, \cite{Reset2}, \cite{Reset3}.  Here, in addition to diffusion, a particle experiences a reset back to some location, and resets form a Poisson process, endowed with a reset rate $r$.  The authors of these sources found there exists an optimal rate, $r^*$ which minimizes the MFPT out of the semi-infinite domain.  We note, however that these sources appear to return the particle back to the reset location once it has hit the absorbing end of the semi-infinite domain, thereby conserving the probability.  This is different from our problem, in which the total probability inside the domain decreases with time, because once particles have reached one of the two absorbing ends, they are not returned back into the domain.

This difference aside, the problem that we are analyzing can be viewed as a version of a reset problem, although the time to reset is not instantaneous.  Moreover, the reset location is not necessarily the location of the attractor $x=X$, since a particle has a chance to return to the diffusive layer before reaching the attractor.  The limit of infinite $v$ would correspond to the instantaneous reset to the attractor, and the limit $b\rightarrow 0$ would cause the resetting to take particles back to $x=X$, i.e. approximating the standard reset problem (although, again, without returning of particles that have reached either of the domain ends).

The dip phenomenon is also observed when $b$ is varied at fixed $a$, see Fig.~\ref{fig:DipComparison_vary_b}.  
At low $b$, MFPT is dominated by the waiting time $1/b$ to return from the attractor to the DL.  A large $b$ asymptote (for $b \gg a$) is the regime of purely diffusive motion - the particles are forced into the DL.  Evidently, having some acccess to the AL leads to a lowering of MFPT because it allows more material to congregate close to one end.

It is interesting to ask what effect increasing the advective velocity would have.  The intuition - supported by the physics of the one-layer model - is that higher $v$ should lead to either an increase of the MFPT or the disappearance of the dip, because with sufficiently large velocity, the density will be more and more localized near the attractor; so, even though the attractor is closer to one end than the other, 
\begin{figure}[h!]
\center \includegraphics[width=3.4in]{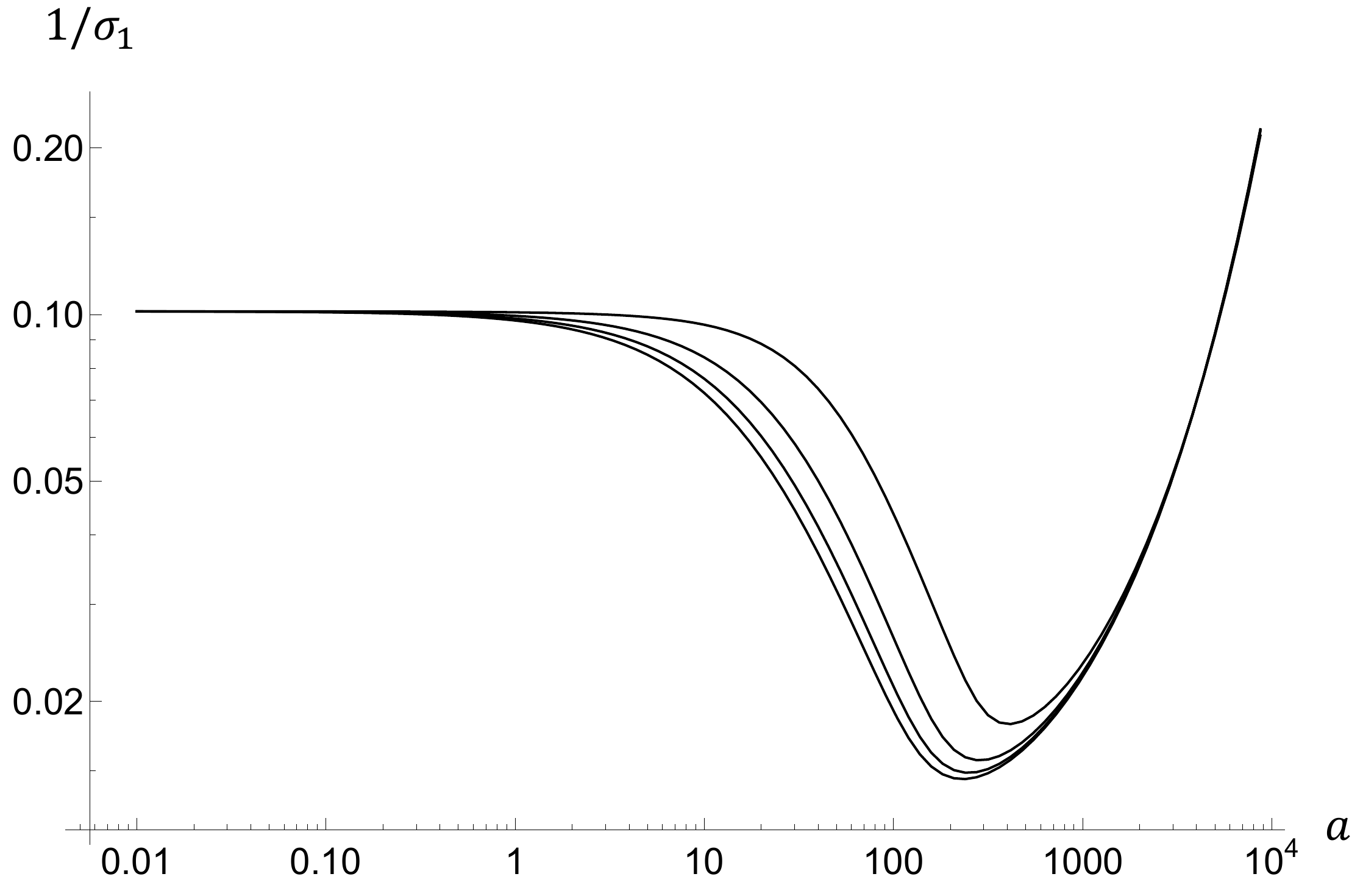}
\caption{Top to bottom: $v=20$, $40$, $60$, $80$.  Here $X=0.5$, and $b=169$.}
\label{fig:engs_dip_multiple_v}
\end{figure}
\begin{figure}[h!]
\center \includegraphics[width=3.4in]{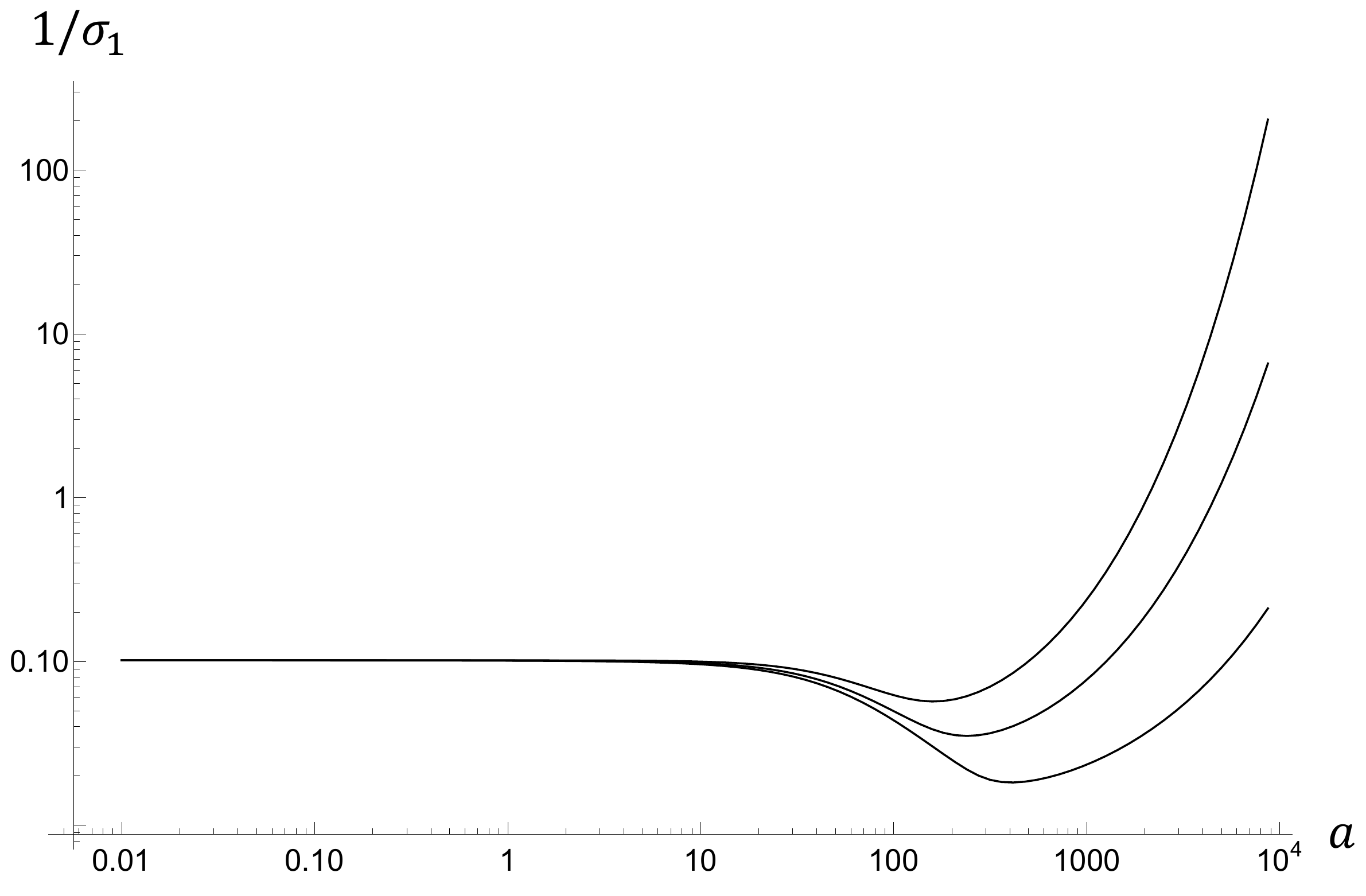}
\caption{Top to bottom: $X=3/26$, $2/26$, $1/26$.  Here $v=20$, and $b=169$.}
\label{fig:engs_dip_multiple_X}
\end{figure}
it is no longer close to this end in comparison to the width of the density distribution.  However, analytical calculations in fact predict the decrease in the value of $1/\sigma_1$ at a fixed $a$ with increasing $v$, see Fig.~\ref{fig:engs_dip_multiple_v}.

An in-depth study of density distributions, which will be published elsewhere \cite{Our_PRE}, sheds light on the reason for this counter-intuitive prediction.  While the density profile in both layers does become more localized with larger velocity (as expected), the part of the profile between the attractor and the close end is not affected;  the decrease in the spread is due to the other side of the profile.  Therefore, as velocity is increased, more and more material is localized near the close end, while the chance of escaping through this end does not diminish - resulting in the overall decrease of escape time.

We also study the effect of varying $X$ in Fig.~\ref{fig:engs_dip_multiple_X}. 
Here the results conform to the intuitive expectation that a decrease in asymmetry will lead to a decrease in the magnitude of the dip (with no dip at all in a completely symmetric geometry).  An attractor placed much closer to the left end than the right one, for example, has two effects. First, it lowers the MFPT overall, since there is less distance to travel during the escape. Second, preventing particles from wandering too far to the right (by increasing $a$, and thus the reset rate) causes the particles to congregate closer to the left end in the more asymmetric situation, leading to a lower MFPT.  

\subsection*{One-layer limit}
\subsubsection*{Dynamics of probability density}
The analytical approach in the one-layer limit is outlined in the Methods section, with details in Appendix B.  
\begin{figure}[h!]
\center \includegraphics[width=3.35in]{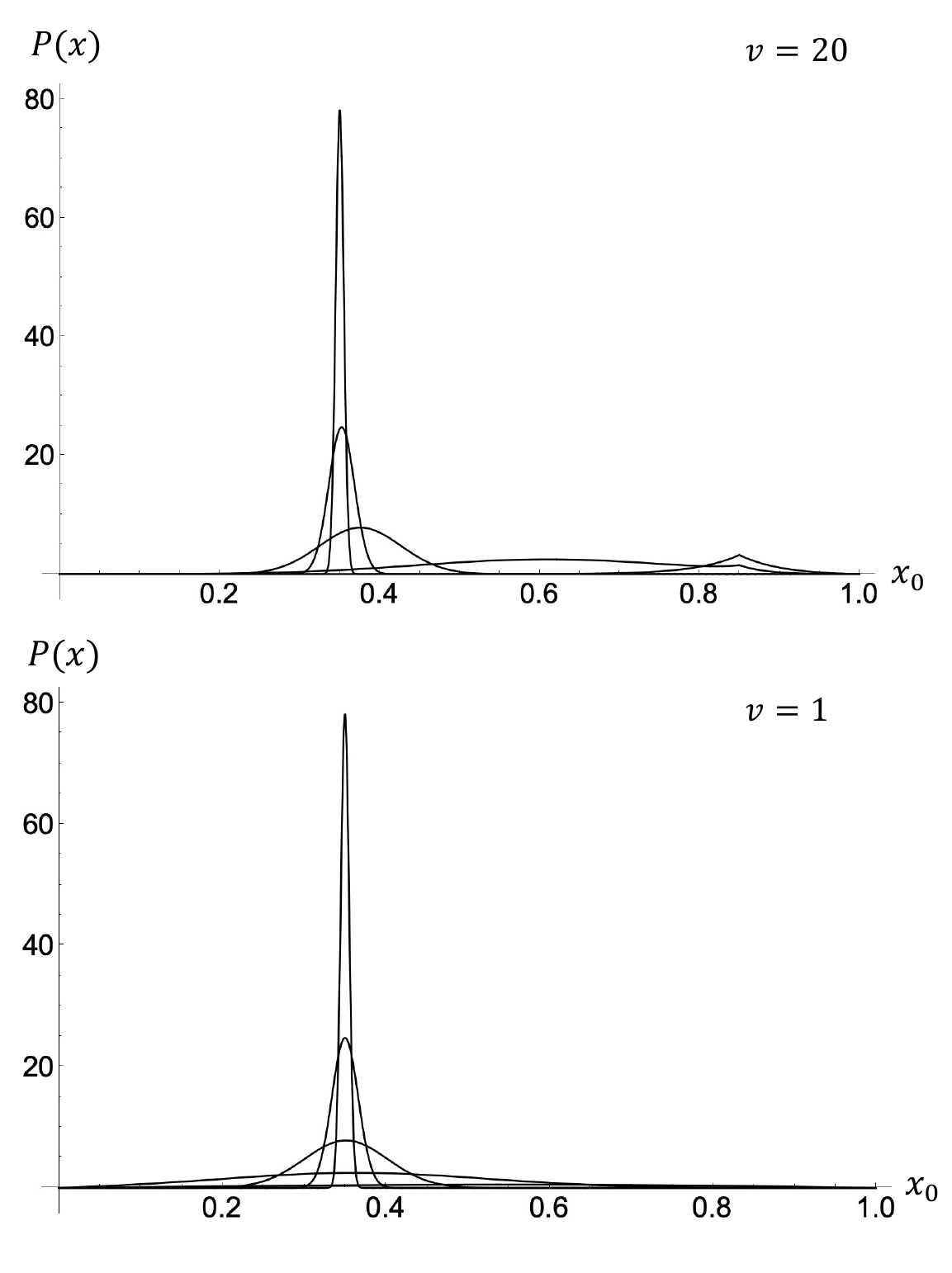}
\caption{$X=0.85$, $x_0=0.35$. The distributions are shown for $t=1.3\times10^{-5}$, $t=1.3\times10^{-4}$, $t=1.3\times10^{-3}$, $t=1.3\times10^{-2}$, $t=1.3\times10^{-1}$.  
Top: $v=20$, bottom: $v=1$.  For $v=1$, the distributions never reach an asymptotic form that is centered on $x_0 = X$. 
}
\label{fig:density_profiles}
\end{figure}
These  predictions are verified by simulations (see Appendix D).  Here we present results of analytical calculations. 

In Fig.~\ref{fig:density_profiles} we show several snapshots in the evolution of the probability density profiles for a specific placement of the attractor and  specific initial condition, for two values of the advective velocity.  Following a $\delta$-function initial condition, there is a quick diffusive spread.  While this spread is happening, the center of the distribution is also advected towards the attractor.  Note that in the $v=1$ case, the average position of particles reaches $1/2$.  On the other hand, for the case of stronger at $v=20$, the center of the distribution reaches the attractor at $x=X$.  

At $v=20$ we begin to see the emergence of large-time asymptotic profile centered on the attractor.  At large times, the distribution reaches a stationary limiting form.  As this profile develops, diffusive spread of the density profile is followed by a contraction, as particles congregate around the attractor and $\sigma_x$ decreases.  At $t\approx 0.06$ the width stops evolving, and the cusp-shaped profile is established in the vicinity of the attractor.  After that, the probability to remain in the domain continues to decrease (the area under the curve will continue to decrease), although the shape of the profile remains stationary.  We will call this limiting profile the large-time distribution or the limiting distribution.  The width of this cusp-shaped limiting distribution decreases with increasing $v$.   At lower $v$, the width also saturates to a constant value at large times, and the limiting distribution also emerges, but it is not centered on the attractor.  

Thus, the picture is this:  the attractor captures some particles and pulls them in to its vicinity at larger $v$, whereas at lower $v$, most of the particles escape before this happens.  The decay rate also decreases - as $v$ grows ever larger, the large-time limiting profile localized around the attractor will decay ever slower, its rate of decay decreasing exponentially with $v$ (this is for sufficiently large $v$, i.e. it is an asymptotic scaling).  In this large $v$ regime, the profile that develops after an initial rapid relaxation may be called quasistationary - as it decays on a time scale much smaller than all other time scales in the problem.  This is the regime of rare events, and we now discuss the scaling of MFPT and escape probabilities in this limiting regime.

\begin{widetext}
\begin{figure}[h]
\center \includegraphics[width=7in]{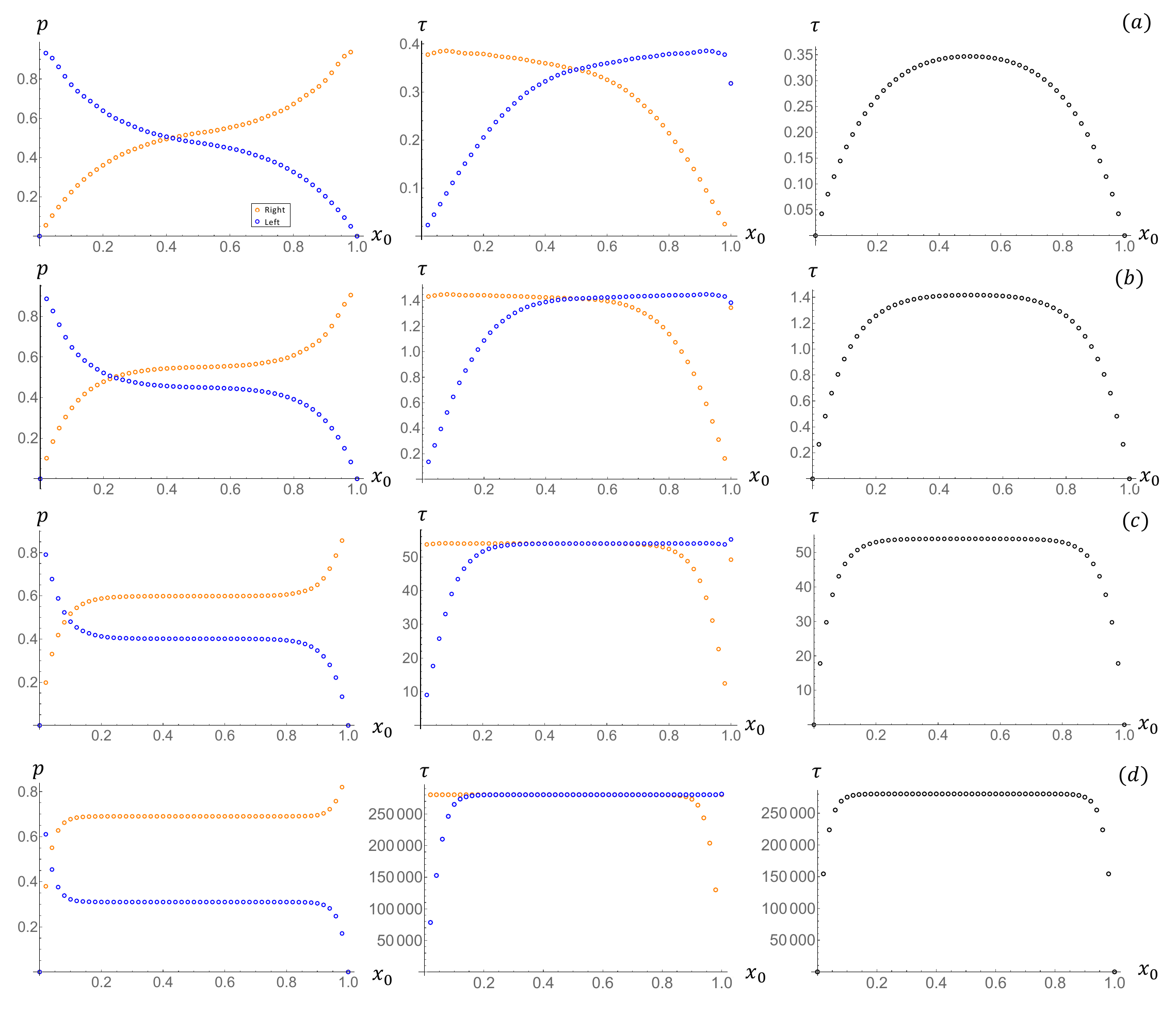}
\caption{Escape probability and MFPT through both ends versus the location $x_0$ of the IC.  The attractor is located at $X=0.51$. (a) $v=5$, (b) $v=10$, (c) $v=20$, (d) $v=40$.  The aberrations at the edge are numerical artifacts. 
}
\label{fig:grid}
\end{figure}
\end{widetext}
\subsubsection*{Scaling of MFPT in the rare event limit}
%

In this regime, various functions of $x_0$ - such as the escape probability and escape time - develop characteristic distinctions between a boundary layer and interior regions.  This is shown in Fig.~\ref{fig:grid}.
As $v$ increases, the MFPT to exit increases, and eventually this time becomes much larger than all the other characteristic time scales of the problem.  In this large $v$ regime, escape becomes a rare event.  Starting from an initial condition $x_0$, a particle will, with overwhelming probability drift towards the fixed point, and fluctuate around it for a time that scales exponentially with $v$ as stated above.  Therefore, the initial condition will be forgotten.  This effect is manifested in Fig.~\ref{fig:grid} by distinct plateaus, that show the absence of dependence on $x_0$.  We show the comparison between such analytical predictions and simulation results of the one-layer regime in Appendix D.  

Escape rates in these plateaus will follow the usual Arrhenius scaling $1/\tau \sim e^{-\Delta U_{eff}/D}$ in physical units.  The effective barrier to escape to the left will be $vX = \Delta U_l$ and the effective barrier to escape to the right will be $v(1-X) = \Delta U_r$.  A small difference between $X$ and $(1-X)$ will be exponentially amplified by large $v$.  Thus, for $0.5< X<1$, the dominant factor will be $v(1-X)$, and therefore, $\tau \sim e^{v(1-X)/D}$, or in dimensionless units, simply 
\begin{equation}
\tau \sim e^{v(1-X)}.
\end{equation}
A more detailed analysis \cite{Our_PRE} predicts the prefactor as well, so the asymptotic expression (i.e. in the rare event regime) is given by $\tau =  4v^{-2} e^{v(1-X)}$.

One comment regarding MFPT results is in order.  We notice that the overall MFPT $\tau$ in Fig.~\ref{fig:grid} is $\approx 2$ times smaller than the $a=b\rightarrow \infty$ limit in Fig.~\ref{fig:NiranjunSym} (see $v=10$ and $v=20$ graphs).  While a small difference is due to slightly different $X$ ($0.51$ in Fig.~\ref{fig:grid} vs.~$0.5$ in Fig.~\ref{fig:NiranjunSym}),  the main reason for this difference is that in the two-layer problem, the advection and diffusion take turns, while they take place simulataneously in the two-layer model.  Thus, all timescales are slowed down by exactly a factor of two in the two-layer model than its truly one-layer equivalent.  In other words, to make the proper comparison, we must multiply the one layer result by $2$ to match the $a=b\rightarrow \infty$ limit of the two-layer model.

\subsubsection*{Small asymmetry leads to a large bias in the exit location}
\begin{figure}[h!]
\center \includegraphics[width=3in]{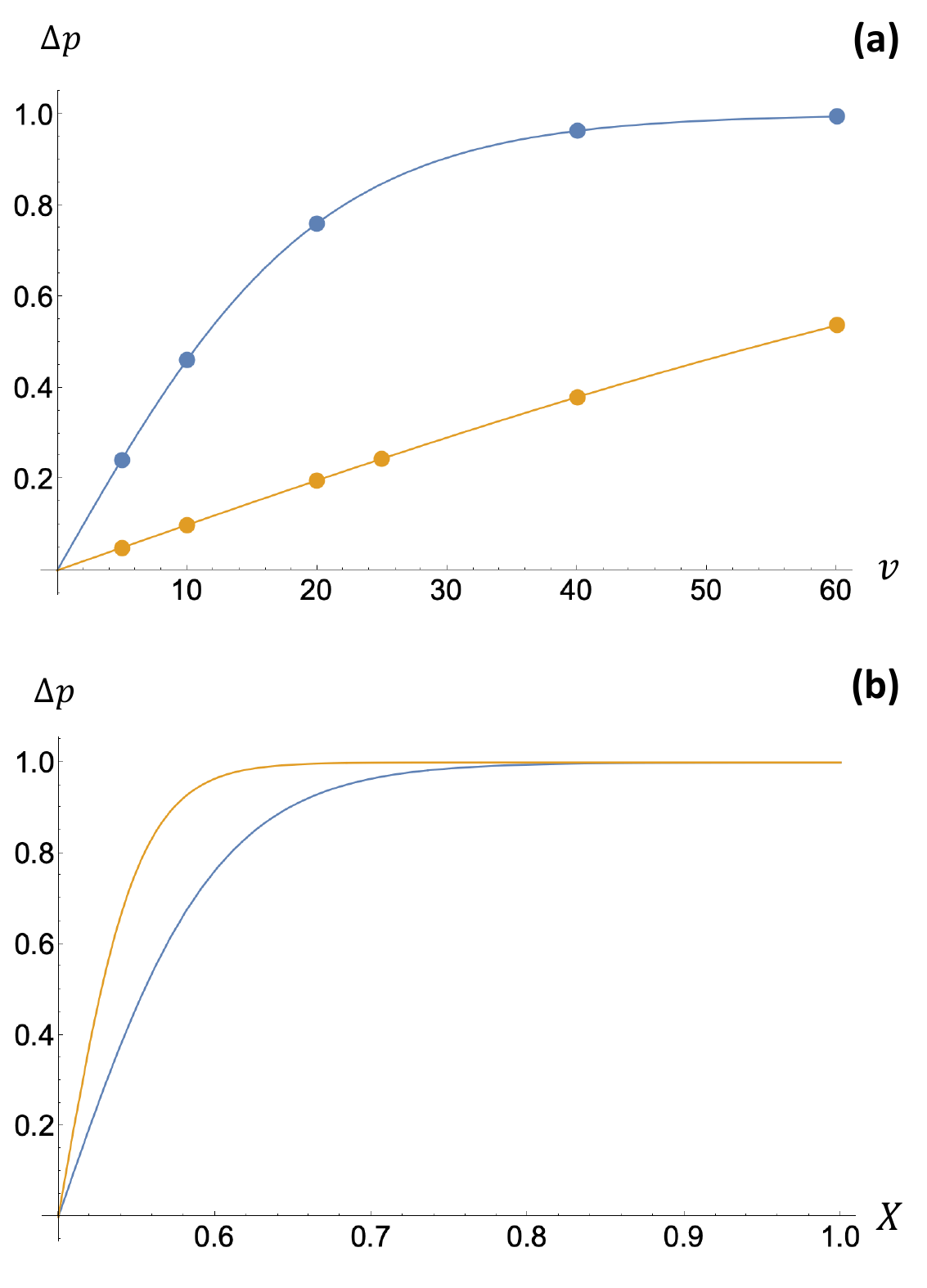}
\caption{(a) $\Delta p$ vs.~$v$.  Top (blue):  $X=0.55$, bottom (orange): $X=0.51$. Dots - full theory, solid curves - Eqn.~(\ref{eq:delta_p}).  (b) $\Delta p$ vs.~$X$, given by  Eqn.~(\ref{eq:delta_p}).  Top (orange): $v=20$, bottom (blue): $v=10$. }
\label{fig:gap_fig}
\end{figure}
One prominent feature of Fig.~\ref{fig:grid} is the amplification in the asymmetry in results (for example $p_l$ and $p_r$ - probabilities to escape through the left and right ends respectively) due to a small asymmetry in the placement of the attractor.  Note that $p_r = ae^{-\Delta U_r}$ and $p_l = ae^{-\Delta U_l}$, where $a$ is some constant.  We can find this constant from the fact that $p_r + p_l = 1$ (a particle definitely exists through one of the two ends eventually).  Thus, $a = \left(e^{-\Delta U_r} + e^{-\Delta U_l}\right)^{-1}$, altogether giving
\begin{equation}
\label{eq:delta_p}
p_r-p_l = \tanh{\left[(X-1/2)v\right]}
\end{equation}
We overlay this prediction on top of $\Delta p$ obtained from the analytic results (depicted in Fig.~\ref{fig:grid}) in Fig.~\ref{fig:gap_fig} (a).
\section*{Conclusion}
In this paper, we looked at a one-dimensional model of intracellular transport via a combination of advection on microtubules and diffusion in the cytoplasm.  This one-dimensional model was motivated by a scenario involving an attractor in the interior of the cell - for example, MTOC.  There are other situations where attractors may arise. Consider, the $\beta$ cell example from the Introduction.  Here motors transport insulin granules along MTs.  Due to orientational disorder \cite{zhu2015microtubules}, several MTs can meet with ends of the same polarity facing each other, forming an aster-like morphological trap (or attractor) for motors that would all congregate at this junction \cite{Ando_Gopinathan}.  It is meaningful to talk about the domain of attraction of such a trap in the following sense.  A molecular motor that attaches to a MT anywhere within this domain will be taken towards the attractor, while a motor that attaches to a mirotubule outside of the domain has a non-zero probability to be taken away from the trap.  When placed inside such a domain - where advective motion along microtubules tends to only attract particles - they can nevertheless escape the domain of attraction of the attractor by desorbing from MTs and diffusing within the cytoplasm until they end up outside of the domain.  Naturally, a question about the time to be stuck in the vicinity of the attractor arises - along with the question of how formation of such traps affects the functioning of the cell and the overall transport of insulin granules across it.  

Using our one-dimensional model, We calculated escape probability through each end, $p_l(x_0)$ and $p_r(x_0)$, and overall $p(x_0)$.  We also calculated the mean first passage time (MFPT) to escape the domain through each end, $\tau_l(x_0)$ and $\tau_r(x_0)$, and overall $\tau(x_0)$.  The initial location inside the cell is determined by the organelles producing the cargo.  The other parameters in the problem were the dimensionless location of the attractor toward which the advective motion is directed, and the dimensionless advective velocity $v$. 

In situations like these, when there is either orientational or polarity disorder, we can think of cells as being divided into domains.

We made several predictions.  When the attractor is placed symmetrically and $a$ and $b$ are finite, there is a crossover between $\tau \sim 0.1$ - diffusive timescale to $\tau$ that grows exponentially in $v$.  The range of $a=b$ over which this crossover happens is wide - a couple of orders of magnitude, but it corresponds to the values of $a$ and $b$ actually found in cells.  This served as our first example of ``fine-tuning'' that allows cells to achieve the biggest change in the functionality with the smallest change in parameter.

For $a=b$ significantly below the crossover, a particle that was released into the diffusive layer has a chance to escape the domain purely diffusively without ever visiting the AL.  For $a=b$ around the crossover value, the probability of this goes to zero - every particle will be advected towards the attractor for at least some of the time.  For $a=b$ significantly above the crossover, the transport enters the effective one-layer regime and exhibits rare event physics.

Asymmetric placement of the attractor gives rise to an interesting phenomenon of an optimal coupling.  Thus, we found that it is possible to minimize the residence time in the domain by increasing the coupling, because that will lower the diffusive spread, and bring particles close to one end of the domain.  

We discussed the effective one-layer regime that results at sufficiently large couplings.  
We also discussed rare event physics that happens at large dimensionless advective velocities.  In such a rare event regime, a portion of particles will be localized in the vicinity of the attractor for a time exponentially long in $v$.  We provide an explicit formula formula for the overall MFPT - including not only the exponential part, but also the prefactor, which scales as $v^{-2}$.  

The idea of exponential sensitivity, and phenomena such as strong amplification of the preferred exit end due to a slight asymmetry is tantalizing.  Extrapolating this finding to two dimensions suggests that in complex, crowded environments that allow for multiple trap-like morphologies (for example, asters), the distribution of cargo around the cell will be non-homogeneous.  This remains to be verified in the future, by extending our model two two dimensions.

Our work is complementary to prior theoretical models of transport that involves a combination of diffusion and advection along microtubules \cite{Previous_theory_1} and \cite{Previous_theory_2}, as neither of these sources are focusing on questions of residence time or the role of asymmetry.

To continue our current work, we would like to study models with reflecting-reflecting or absorbing-reflecting boundary conditions, or models in which the source is on one end and the target is on the other.  Such models would be better suited for 
transport of cargo in cilia \cite{Cilia}, transport between the plasma membrane and Golgi apparatus \cite{Golgi1}, \cite{Golgi2}, or between Endoplasmic Reticulum and Golgi \cite{Golgi3}, \cite{Koslover_review}, transport of viruses towards replication sites \cite{Virus1}, \cite{Virus2}, and other intracellular transport situations \cite{Koslover_review}, \cite{Interactome}.

This work was supported by the National Science Foundation (NSF-DMS-1616926 to AG) and NSF-CREST: Center for Cellular and Bio-molecular Machines at UC Merced (NSF-HRD-1547848 and 2112675 to AG). AG and NS also acknowledge partial support from the NSF Center for Engineering Mechanobiology grant CMMI-154857 and computing time on the Multi-Environment Computer for Exploration and Discovery (MERCED) cluster at UC Merced (NSF-ACI-1429783). NS acknowledges Graduate Student Opportunity Program Fellowship from the University of California, Merced. BR acknowledges the support of the William and Linda Cal Poly Frost fund for undergraduate research.

\newpage
\begin{appendix}
\begin{widetext}
\section*{A: Details of the two-layer calculations}
We start with the full one-dimensional, two-layer model in dimensionless form (primes have been omitted for clarity):
\begin{eqnarray}
\label{eq:RhoEq2} \frac{\partial \rho}{\partial t} &=& -\frac{\partial}{\partial x}\left(v(x)\rho\right) + a \theta - b\rho \\
\label{eq:ThetaEq2}\frac{\partial \theta}{\partial t} &=& -a\theta + b\rho +  \frac{\partial^2 \theta}{\partial x^2}
\end{eqnarray}
Here $a$ and $b$ are respectively the rates of adsorption to and desorption from microtubules, $v$ is the dimensionless velocity profile, $\rho$ is the density of particles on microtubules, and $\theta$ is the density of particles diffusing in the cytoplasm.

We seek modal solutions (or eigensolutions) of the form 
\begin{eqnarray}
\left(\begin{array}{c} \rho \\ \theta \end{array}\right) = \left(\begin{array}{c} R(x) \\ \Theta(x) \end{array}\right)e^{-\sigma t}.
\end{eqnarray}
The vector $\left(\begin{array}{c} R(x) \\ \Theta(x) \end{array}\right)$ is an eigenvector of the operator (see Eq.~(11) of the text) that represents minus the right hand side of Eqs.~(\ref{eq:RhoEq2})-(\ref{eq:ThetaEq2}), and $\sigma$ is an eigenvalue of this operator.

However, due to the mass accumulation at the attractor, we must also include a $\delta$-function term to accommodate for this mathematically.  The mass will not accumulate at the junction point  due to the diffusive term that acts on the diffusive layer density.  Note also that the $\delta$- function in the advective layer acts like a point source for the diffusive layer. When we study a simple diffusive problem with a $\delta$-function source plus absorbing boundaries, and seek a steady-state (time independent) solution, the density profile does not acquire a $\delta$-function response - the diffusion acts infinitely quickly to dissipate such a singularity.   With this in mind, we must augment the above formula to 
\begin{equation}
\label{eq:augment}
\left(\begin{array}{c} \rho \\ \theta \end{array}\right) = \left(\begin{array}{c} R(x) \\ \Theta(x) \end{array}\right)e^{-\sigma t}  + \left(\begin{array}{c} 1 \\ 0\end{array}\right) \left(M_0  e^{-\sigma t}\right) \delta(x-X).
\end{equation}
Substituting this back to Eqs.~(\ref{eq:RhoEq2})-(\ref{eq:ThetaEq2}), and setting $Q = \frac{d\Theta}{dx}$, we get 
\begin{equation}
\label{eq:left_matrix}
\frac{d}{dx} \left(\begin{array}{c} R \\ Q \\ \Theta \end{array}\right) = \left(\begin{array}{ccc} (-\frac{b}{v} + \frac{\sigma}{v}) & 0 & \frac{a}{v} \\ -b & 0 & (a - \sigma) \\ 0 & 1 & 0 \end{array}\right)  \left(\begin{array}{c} R \\ Q \\ \Theta \end{array}\right)
\end{equation}
for $0 \leq x<X$ (call it Region-I) and 
\begin{equation}
\label{eq:right_matrix}
\frac{d}{dx} \left(\begin{array}{c} R \\ Q \\ \Theta \end{array}\right) = \left(\begin{array}{ccc} (\frac{b}{v} - \frac{\sigma}{v}) & 0 & -\frac{a}{v} \\ -b & 0 & (a - \sigma) \\ 0 & 1 & 0 \end{array}\right)  \left(\begin{array}{c} R \\ Q \\ \Theta \end{array}\right),
\end{equation}
for $X<x\leq  1$ (call it Region-II). 
The solutions, will take the form:
 \begin{equation}
 \label{eq:solnI}
\left(\begin{array}{c} R_I \\ Q_I \\ \Theta_I \end{array}\right) = A \left(\begin{array}{c} v^1_R \\ v^1_Q \\ v^1_\Theta \end{array}\right) e^{\lambda_1 x} + B \left(\begin{array}{c} v^2_R \\ v^1_Q \\ v^2_\Theta \end{array}\right) e^{\lambda_2 x} + C \left(\begin{array}{c} v^3_R \\ v^3_Q \\ v^3_\Theta \end{array}\right) e^{\lambda_3 x}
 \end {equation}
in Region-I and
\begin{equation}
\label{eq:solnII}
\left(\begin{array}{c} R_{II} \\ Q_{II} \\ \Theta_{II} \end{array}\right) = D \left(\begin{array}{c} w^1_R \\ w^1_Q \\ w^1_\Theta \end{array}\right) e^{\mu_1 x} + E \left(\begin{array}{c} w^2_R \\ w^1_Q \\ w^2_\Theta \end{array}\right) e^{\mu_2 x} + F \left(\begin{array}{c} w^3_R \\ w^3_Q \\ w^3_\Theta \end{array}\right) e^{\mu_3 x},
 \end {equation}
 in Region-II.
 The $\vec{v}$s and $\lambda$s are eigenvectors and eigenvalues of the matrix in Eq.~(\ref{eq:left_matrix}), while $\vec{w}$s and $\mu$s are eigenvectors and eigenvalues of the matrix in Eq.~(\ref{eq:right_matrix}).  The $\lambda$s satisfy the equation
\begin{equation}
-\lambda^3 + \left(\frac{\sigma - b}{v}\right) \lambda^2 + (a-\sigma ) \lambda + \frac{\sigma^2 -\sigma (a+b)}{v} = 0,
\end{equation}
 and the $\mu$s satisfy the equation
 \begin{equation}
-\mu^3 - \left(\frac{\sigma-b}{v}\right) \mu^2 + (a-\sigma ) \mu - \frac{\sigma^2 -\sigma (a + b)}{v} = 0.
\end{equation}
The eigenvectors have the structure
\begin{equation}
\vec v = \left(\begin{array}{c} \frac{-\lambda^2 + a - \sigma}{b} \\ \lambda \\ 1 \end{array}\right),
\end{equation}
and 
\begin{equation}
\vec w = \left(\begin{array}{c} \frac{-\mu^2 + a - \sigma}{b} \\ \mu \\ 1 \end{array}\right).
\end{equation}
The functions on both sides of the attractor are different, and they need to be stitched correctly.  The stitching is determined by the boundary conditions, so we now discuss these.  The boundary conditions will determine the eigenvalues $\sigma_n$.  We note that there are seven unknowns: coefficients A - F (see Eqs.~(\ref{eq:solnI})-(\ref{eq:solnII})), and the mass growth rate $M_0$ (see Eq.~(\ref{eq:augment}), so we need seven constraints (or conditions).

First, there are absorbing boundary conditions at each end, which require that $R(x=0)=\Theta(x=0)=0$ and $R(x=1)=\Theta(x=1)=0$.  The additional three conditions come from the location of stitching, i.e. the attractor location at $x=X$.  The diffusive layer density must be continuous to avoid infinite currents.  Thus, $\Theta_I(X) = \Theta_{II}(X)$.  The remaining two boundary conditions come from mass conservation.  To extract these, we integrate Eqs.~(\ref{eq:RhoEq2})-(\ref{eq:ThetaEq2}) through the junction point, i.e. from $x-\epsilon$ to $x+\epsilon$ for arbitrarily small $\epsilon$.  Performing this on Eq.~(\ref{eq:RhoEq2}) gives
\begin{equation}
-\sigma M_0  = -bM_0 - \left(v_{II}R_{II}(X) - v_IR_{I}(X)\right) = -bM_0 + v\left(R_{II}(X) +  R_I(X)\right). 
\end{equation}
Note that the temporal terms would not be absent if the $\delta$-function component of $\rho$ was not proportional to $e^{-\sigma t}$.  This equation says that the rate of growth of the advective layer mass at $x=X$ (i.e. of the strength of the $\delta$-function) is driven by the inflow from this layer, and outflow into the diffusive layer.  Performing the integration on Eq.~(\ref{eq:ThetaEq}) gives
\begin{equation}
bM_0   = \left. \frac{d\Theta_I}{dx}\right|_{x=X} - \left. \frac{d\Theta_{II}}{dx}\right|_{x=X}.
\end{equation}
This equation says that any difference in the outflow rates (i.e. different slopes of the diffusive layer density) is balanced by the inflow from the advective layer.

We now implement these boundary conditions algebraically.  
We have:
\begin{enumerate}
	\item Absorbing boundary condition at $x=0$ in the advective layer: $R_{I} (x = 0) = 0$
	\begin{equation}
	A \left(\frac{-\lambda_1^2 + a - \sigma}{b}\right) + B \left(\frac{-\lambda_2^2 + a - \sigma}{b}\right) + C \left(\frac{-\lambda_3^2 + a - \sigma}{b}\right) = 0
	\end{equation}

	\item Absorbing boundary condition at $x=0$ in the diffusive layer: $\Theta_{I} (x = 0) = 0$
	\begin{equation}
	A + B + C = 0 
	\end{equation}

	\item Absorbing boundary condition at $x=1$ in the advective layer: $R_{II} (x = 1) = 0$
	\begin{equation}
	D \left(\frac{-\mu_1^2 + a - \sigma}{b}\right)  e^{\mu_1} + E \left(\frac{-\mu_2^2 + a - \sigma}{b}\right) e^{\mu_2} + F \left(\frac{-\mu_3^2 + a - \sigma}{b}\right) e^{\mu_3} = 0 
	\end{equation}

	\item Absorbing boundary condition at $x=1$ in the diffusive layer: $\Theta_{II} (x = 1) = 0$
	\begin{equation}
	D e^{\mu_1}  + E e^{\mu_2} +F e^{\mu_3} = 0 
	\end{equation}

	\item Continuity at $x=X$ in the diffusive layer (to prevent infinite diffusive currents): $\Theta_{I} (x = X) = \Theta_{II} (x = X)$
	\begin{equation}
	A e^{\lambda_1 X} + B e^{\lambda_2 X} + C e^{\lambda_3 X} = D e^{\mu_1 X} +E e^{\mu_2 X} + F e^{\mu_3 X}
	\end{equation}

	\item Mass conserving boundary condition in advective layer: $R_{II} (x = X) + R_{I} (x = X)= \frac{b - \sigma}{v} M_0$
	\begin{eqnarray}
	&& D \left(\frac{-\mu_1^2 + a - \sigma}{b}\right) e^{\mu_1 X} + E \left(\frac{-\mu_2^2 + a - \sigma}{b}\right) e^{\mu_2 X} + F \left(\frac{-\mu_3^2 + a - \sigma}{b}\right) e^{\mu_3 X} \nonumber \\ 
	&+&  A \left(\frac{-\lambda_1^2 + a - \sigma}{b}\right) e^{\lambda_1 X} + B \left(\frac{-\lambda_2^2 + a - \sigma}{b}\right) e^{\lambda_2 X} + C \left(\frac{-\lambda_3^2 + a - \sigma}{b}\right) e^{\lambda_3 X} = \frac{b-\sigma}{v} M_0  \nonumber \\
	\end{eqnarray}

	\item Mass conserving boundary condition in diffusive layer: $\left.\frac{d\Theta_I}{dx}\right|_{x = X} - \left.\frac{d\Theta_{II}}{dx} \right|_{x = X} = b M_0$ 
	\begin{equation}
	A \lambda_1 e^{\lambda_1 X} + B \lambda_2 e^{\lambda_2 X} + C \lambda_3 e^{\lambda_3 X} - D \mu_1 e^{\mu_1 X} - E \mu_2 e^{\mu_2 X} - F \mu_3 e^{\mu_3 X} = b M_0.
	\end{equation}
\end{enumerate}
We can write all these seven equations in the compact matrix form:
\scriptsize
\begin{eqnarray}
\label{eq:BigM}
\left(\begin{array}{ccccccc} \frac{-\lambda_1^2 + a - \sigma}{b} & \frac{-\lambda_2^2 + a - \sigma}{b} & \frac{-\lambda_3^2 + a - \sigma}{b} & 0 & 0 & 0 & 0 \\ 1 & 1 & 1 & 0 & 0 & 0 & 0 \\ 0 & 0 & 0 & \frac{-\mu_1^2 + a - \sigma}{b}  e^{\mu_1} & \frac{-\mu_2^2 + a - \sigma}{b} e^{\mu_2} & \frac{-\mu_3^2 + a - \sigma}{b} e^{\mu_3} & 0 \\ 0 & 0 & 0 & e^{\mu_1} & e^{\mu_2} & e^{\mu_3} & 0 \\ e^{\lambda_1 X} & e^{\lambda_2 X} & e^{\lambda_3 X} & -e^{\mu_1 X} & -e^{\mu_2 X} & -e^{\mu_3 X} & 0 \\ \frac{-\lambda_1^2 + a - \sigma}{b} e^{\lambda_1 X} & \frac{-\lambda_2^2 + a - \sigma}{b} e^{\lambda_2 X} & \frac{-\lambda_3^2 + a - \sigma}{b} e^{\lambda_3 X} & \frac{-\mu_1^2 + a - \sigma}{b} e^{\mu_1 X} & \frac{-\mu_2^2 + a - \sigma}{b} e^{\mu_2 X} & \frac{-\mu_3^2 + a - \sigma}{b} e^{\mu_3 X} & \frac{\sigma-b}{v} \\ \lambda_1 e^{\lambda_1 X} & \lambda_2 e^{\lambda_2 X} & \lambda_3 e^{\lambda_3 X} & -\mu_1 e^{\mu_1 X} & -\mu_2 e^{\mu_2 X} & - \mu_3 e^{\mu_3 X} & -b \end{array}\right)  
\left(\begin{array}{c} A \\ B \\ C \\ D \\ E \\ F \\ M_0 \end{array}\right) = \left(\begin{array}{c} 0 \\ 0 \\ 0 \\ 0 \\ 0 \\ 0 \\ 0 \end{array} \right). \nonumber \\
\end{eqnarray}
\normalsize
Because of the structure of this equation, we see that (i) the determinant must be non-zero for a non-trivial solution and (ii) the nontrivial solution is non-unique - it has at least one degree of freedom. For instance, we are free to choose one of the variables, or we are free to choose the normalization.  Making use of this freedom, we chose to set $M_0=1$.  These equations were then used to solve for the remaining coefficients $A$, $B$, $C$, $D$, $E$, and $F$.   

Thus, calling the matrix involved in Eq.~(\ref{eq:BigM}), M, $\text{Det}(M)=0$ should provide an algebraic equation for $\sigma$.

Expanding determinant in terms of minors, we have
\begin{equation}
0 = b \text{Det}\left(m_{77}\right)  +\left(\frac{\sigma-b}{v}\right) \text{Det}\left(m_{67}\right)
\end{equation}
where the minor $m_{ij}$ is a matrix obtained by removing $i$th row and $j$the column from $M$.

Once $M_0$ is chosen, the coefficients $(A,...,F)$ must be unique.  This means that both $\text{Det}\left(m_{77}\right)$ and $\text{Det}\left(m_{67}\right)$ must both be non-zero.  If $\text{Det}\left(m_{77}\right)$ is non-zero, then the solution $(A, ... ,F)$ obtained from the first six equations can be found with the inverse of $m_{77}$, and is unique.  This implies that $\text{Det}\left(m_{67}\right)$ must also be non-zero (otherwise, the solution $(A, ... ,F)$ obtained from the first five and the seventh equation is non-unique, leading to a contradiction).

Thus, the kind of a zero of $\text{Det}(M)$ that we want is one in which $\text{Det}\left(m_{77}\right)$ and $\text{Det}\left(m_{67}\right)$ are both non-zero.  Therefore, we're interested in the zeros of the following quantity:
\begin{equation}
\text{Det}' = b  + \left(\frac{\sigma-b}{v}\right) \frac{\text{Det}\left(m_{67}\right)}{\text{Det}\left(m_{77}\right)}.
\end{equation}
It is the zeros of this determinant that gives us $\sigma$ in terms of $(a,b,v,X)$.  

We were primarily interested in the lowest (ground state) eigenvalue $\sigma_1$, and the inverse $1/\sigma_1$ that serves as a characteristic measure of the escape time\footnote{This is especially true in the rare event regime that develops at sufficiently large $v$ - when $\sigma_1$ should be separated from the rest of $\sigma$s by a gap that grows exponentially in $v$ - while there is no such gap between the rest of the eigenvalues.}.  Because the set of eigenfunctions and eigenvalues turned out to be finite, they are of limited value in being able to construct a solution that fits the $\delta$-function initial condition, and thereby to properly compute MFPT.

\newpage
\section*{B: One-layer theory}
\label{sec:OneLayer}
We now discuss the computation of the eigenfunctions $p(x)$.  The subscript $n$ will be dropped to lighten the notation.  Recall that  $0<x<X$ is Region-I, and that $X<x<1$ is Region-II.  The eignfunctions satisfy 
\begin{equation}
\label{eq:Oleft}
\sigma p = -v\frac{dp}{dx} + \frac{d^2p}{dx^2}
\end{equation}
in Region-I, and 
\begin{equation}
\label{eq:Oright}
\sigma p = v\frac{dp}{dx} + \frac{d^2p}{dx^2}
\end{equation}
in Region-II.  The solution in Region-I is $p_I=a_I e^{\lambda_+ x} + b_I e^{\lambda_- x}$, where the $\lambda$s satisfy
\begin{equation}
\label{eq:lambdas}
\lambda_{\pm} = \frac{v\pm\sqrt{v^2+4\sigma}}{2}.
\end{equation}
The solution in Region-II is $p_{II} = a_{II} e^{\mu_+ x} + b_{II} e^{\mu_- x}$, where the $\mu$s satisfy
\begin{equation}
\label{eq:mus}
\mu_{\pm} = \frac{-v\pm\sqrt{v^2+4\sigma}}{2}.
\end{equation}
The coefficients $a$ and $b$ will be fixed with the following four boundary conditions (BCs).  The first two are the absorbing BCs at the ends, $p_I(0)=p_{II}(1)=0$.  The third boundary condition is the continuity of the solution $p_I(X) = p_{II}(X)$.  A discontinuous solution is unphysical due to the diffusion term
.  In a one-layer theory, there will not be an accumulation of mass at the trap, i.e. there will be no term like $\delta(x-X)$.  Any such density would be immediately smoothed out by the action of the diffusion.  Note that in the full, two-layer theory, such term existed only in the advective layer, but not in the diffusive layer.  In the absence of a $\delta$-function-like accumulation of mass at $x=X$, the currents across $x=X$ will be continuous.  This gives us the fourth boundary condition that enforces the continuity of currents at the junction: $vp_I(X) - \left.\frac{dp_I}{dx}\right|_{x=X} = -vp_{II}(X) - \left.\frac{dp_{II}}{dx}\right|_{x=X}$.  

Applying these four boundary conditions leads to four equations:
\footnotesize
\begin{eqnarray}
a_I + b_I &=& 0 \\
a_{II} e^{\mu_+} + b_{II}e^{\mu_-} &=& 0 \\
a_Ie^{\lambda_+ X} + b_Ie^{\lambda_- X} &=& a_{II}e^{\mu_+ X} + b_{II}e^{\mu_-X} \\
v\left(a_Ie^{\lambda_+ X} + b_Ie^{\lambda_- X}\right) - \left(\lambda_+ a_Ie^{\lambda_+ X} + \lambda_- b_Ie^{\lambda_- X}\right) &=& -v\left(a_{II}e^{\mu_+ X} + b_{II}e^{\mu_- X}\right) - \left(\mu_+ a_{II}e^{\mu_+ X} + \mu_- b_{II}e^{\mu_- X}\right) \nonumber  \\
\end{eqnarray}
\normalsize
Substituting the first two into the last two gives
\footnotesize
\begin{eqnarray*}
a_I\left(e^{\lambda_+ X} - e^{\lambda_- X}\right) &=& a_{II}\left(e^{\mu_+ X} - e^{\mu_+ - \mu_-} e^{\mu_-X}\right) \\
va_I\left(e^{\lambda_+ X} - e^{\lambda_- X}\right) - a_I\left(\lambda_+ e^{\lambda_+ X} - \lambda_- e^{\lambda_- X}\right) &=& -va_{II}\left(e^{\mu_+ X} - e^{\mu_+ - \mu_-} e^{\mu_- X}\right) - a_{II} \left(\mu_+e^{\mu_+ X} - \mu_- e^{\mu_+ - \mu_-} e^{\mu_- X}\right)
\end{eqnarray*}
\normalsize
Using the first of these, and substituting into the second we obtain
\scriptsize
\begin{equation}
\label{eq:eigenvalue_equation}
v\left(e^{\lambda_+ X} - e^{\lambda_- X}\right) - \left(\lambda_+ e^{\lambda_+ X} - \lambda_- e^{\lambda_- X}\right) - \left[-v\left(e^{\mu_+ X} - e^{\mu_+ - \mu_-} e^{\mu_- X}\right) -  \left(\mu_+e^{\mu_+ X} - \mu_- e^{\mu_+ - \mu_-} e^{\mu_- X}\right)\right]\left(\frac{e^{\lambda_+ X} - e^{\lambda_- X}}{e^{\mu_+ X} - e^{\mu_+ - \mu_-} e^{\mu_-X}}\right) = 0,
\end{equation}
\normalsize
where $\lambda$s and $\mu$s are given by Eqs.~(\ref{eq:lambdas}) and (\ref{eq:mus}) respectively.  Eq.~(\ref{eq:eigenvalue_equation}) is an equation for eigenvalues $\sigma$ as a function of $v$ and $X$.  Moreover, 
\begin{equation}
\label{eq:pI}
p_I = \left(e^{\lambda_+ x} - e^{\lambda_- x}\right),
\end{equation}
and 
\begin{equation}
\label{eq:pII}
p_{II} = \left(\frac{e^{\lambda_+ X} - e^{\lambda_- X}}{e^{\mu_+ X} - e^{\mu_+ - \mu_-}e^{\mu_- X}}\right)\left(e^{\mu_+ x} - e^{\mu_+ - \mu_-} e^{\mu_- x}\right)
\end{equation}
The modes given this way are not normalized; they will be normalized below.  We will see below that eigenvalues turn out to be real.  

The coefficients $c_n$ are determined as usual by the initial condition, $P(x,t=0) = \sum_n c_n p_n(x)$.  Because the operator $O$ is non-Hermitian, eigenfunctions are generally non-orthogonal, i.e. $\int_0^1 p^*_n(x)p_m(x)\,dx \neq 0$, so we can't compute $c_m$ with the help of an inner product $\int_0^1 P(x,0)p_m(x)\,dx$.  However, eigenfunctions of the adjoint operator $O^{\dag}$ have the property that they are either orthogonal to the eigenfunctions of $O$, or otherwise have eigenvalues that are complex conjugates of each other.

Therefore, in order to be able to express initial conditions, we need to compute a set of eigenfunctions and eigenvalues of $O^{\dag}$.  Even after this, there is no guarantee that we will be able to express any initial condition, because there's also no guarantee of completeness, due to operators being non-Hermitian.

The adjoint of $O$ is given by 
\begin{equation}
O^{\dag} = v(x) \frac{d}{dx} + \frac{d^2}{dx^2}.
\end{equation}
To find the eigenfunctions of this operator, it helps to look back at the original equation with operator $O$.  We note that both Eqs.~(\ref{eq:Oleft})-(\ref{eq:Oright}) can be written compactly as one equation
\begin{equation}
\sigma p = \frac{d}{dx}\left(\frac{dU}{dx}p + \frac{d^2 p}{dx^2}\right),
\end{equation}
where the potential (in analogy with physics) $U$ is given by
\begin{equation}
\label{eq:U}
U(x) = \left\{\begin{array}{cc} v(X-x) & x \leq X, \\
v(x-X) & x \geq X,
\end{array} \right.
\end{equation}
or, more compactly, $v(x) = -\frac{dU}{dx}$.

Now, let $p=q(x)e^{-U(x)}$ - we can always do this.  Substituting this ansatz we find that  $q(x)$ obeys 
\begin{eqnarray}
\sigma_1 q &=& -\frac{dU}{dx}\frac{dq}{dx} + \frac{d^2 q}{dx^2} \nonumber \\
&=& v(x)\frac{dq}{dx} + \frac{d^2 q}{dx^2}.
\end{eqnarray}
That is, $q = p(x)e^{U(x)}$ is the eigenfunction of the adjoint operator that we were seeking!  Moreover, it has the same eigenvalue as the operator $O$.  The modes given this way are not normalized; they will be normalized below. 

For operators with a finite dimensional eigenspace, eigenvalues of an adjoint operator $O^\dag$ are  complex conjugates of the eigenvalues of the operator $O$.  In such cases, equality of the two sets of eigenvalues implies that they are real.  In our case the eigenspace is not guaranteed to be finite (in fact, we hope that it isn't, if there is any chance at completeness).  However, our numerical investigation revealed that eigenvalues $\sigma$ are always real (and negative). 

Next, we give an example of the result of several hundred low-lying eigenvalues.  The first example is for $X=0.85$ and $v=1$.
\begin{figure}[h!]
\center \includegraphics[width=4in]{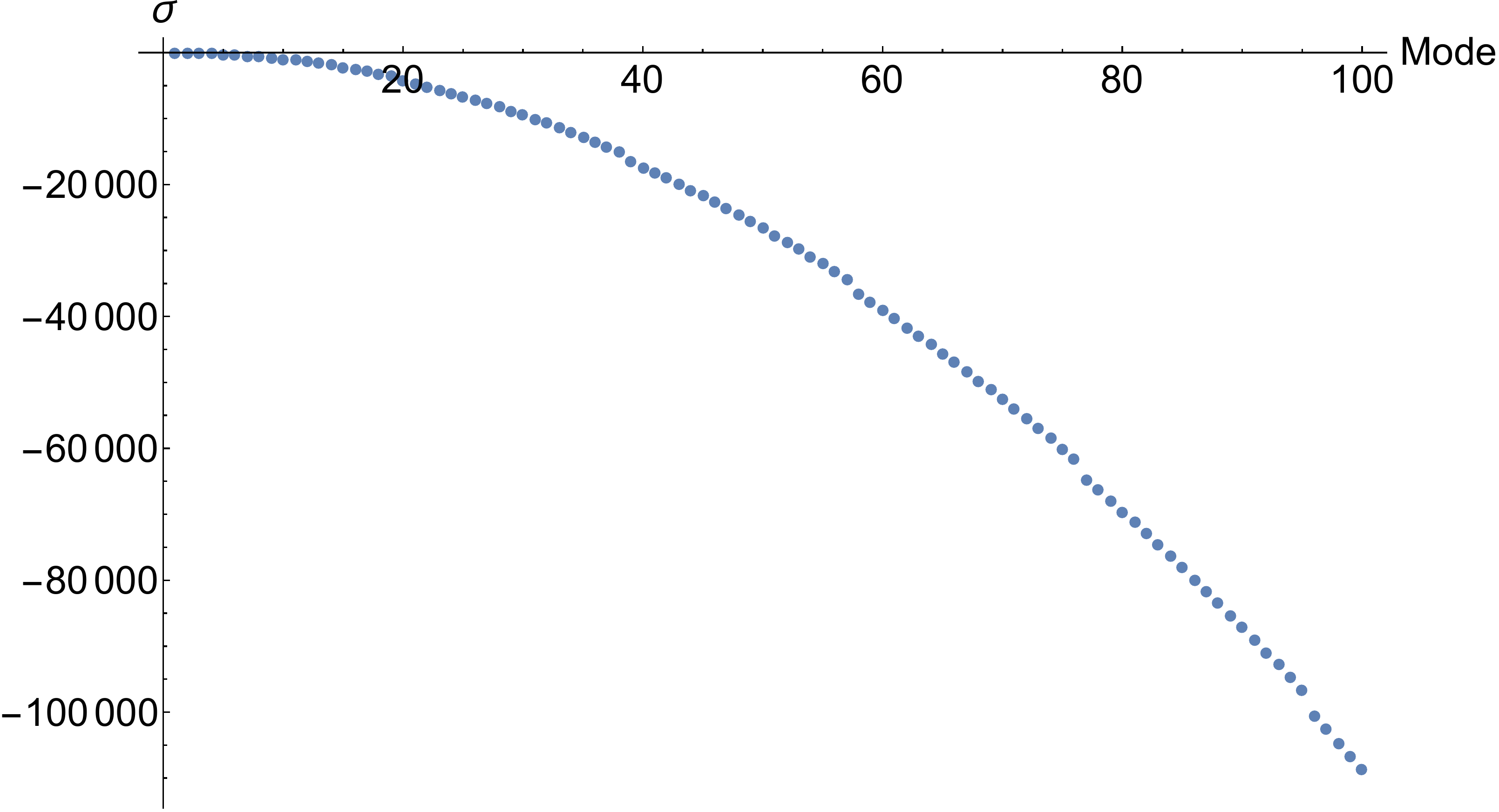}
\caption{Lowest $100$ eigenvalues for $X=0.85$ and $v=1$.}
\end{figure}
We observe an interesting feature that eigenvalues appear in groups.  The second example is for $X=0.6$ and $v=1$. 
\begin{figure}[h!]
\center \includegraphics[width=4in]{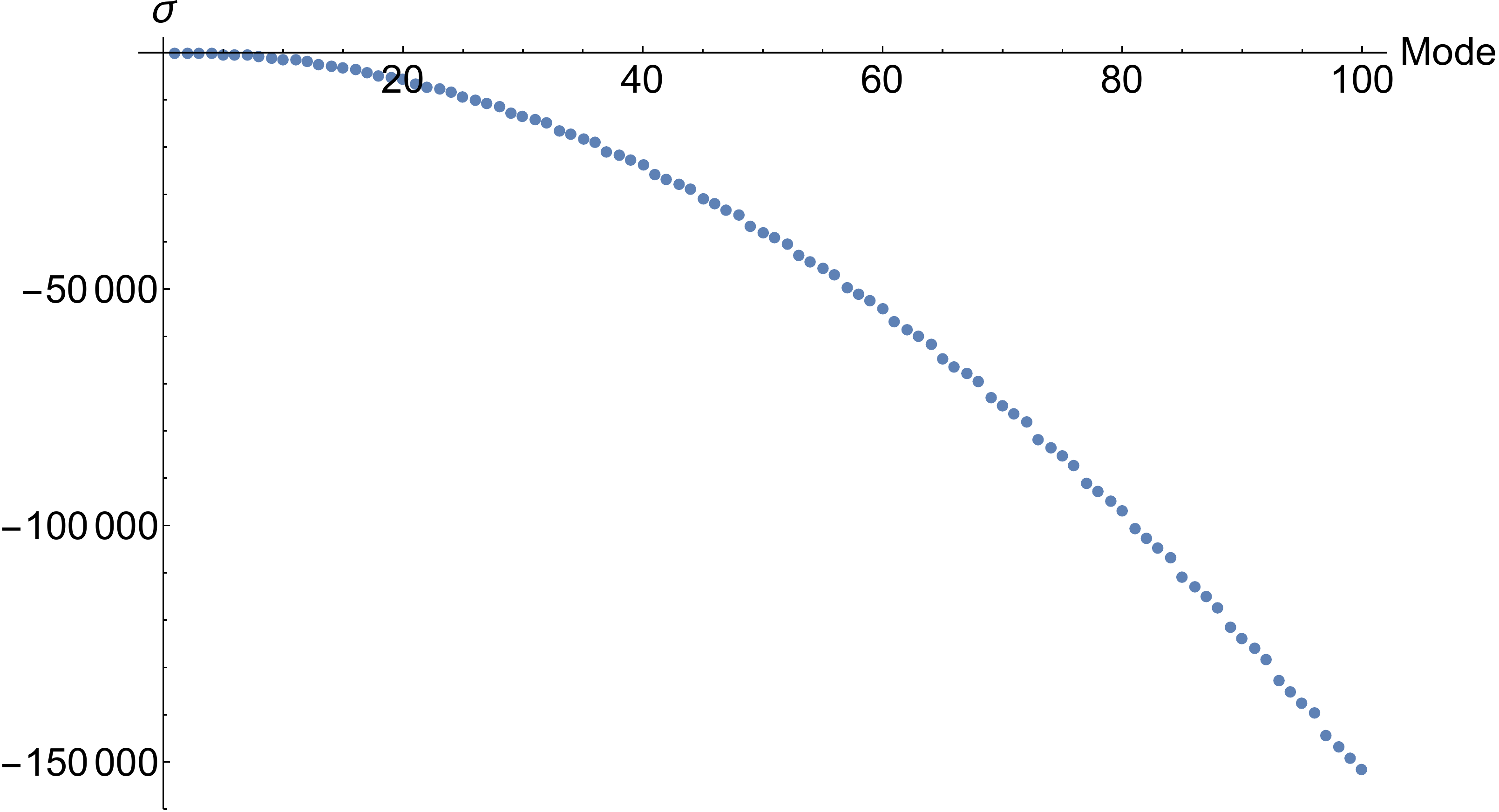}
\caption{Lowest $100$ eigenvalues for $X=0.6$ and $v=1$.}
\end{figure}
We notice that the size of groups has changed.  There is no obvious relation between $X$ the the size of groups - for instance, for $X=0.55$ the groups are again increased in size.

We verified numerically the orthogonality of several eigenfunctions belonging to different eigenvalues, and found it to hold true. Eigenfunctions were also normalized by multiplying by the following factors: 
\begin{eqnarray*}
a_p &=& \frac{1}{\sqrt{\int_0^1 p_n^*(x)p_n(x)}}, \\ 
A_q &=& \frac{1}{\sqrt{\int_0^1 q_n^*(x)q_n(x)}}, 
\end{eqnarray*}
where $p$s and $q$s are given by Eqs.~(\ref{eq:pI})-(\ref{eq:pII}) and $q = p(x)e^{U(x)}$.  The resultant modes came out to be either purely real or purely imaginary.  In this latter case, they can be made real by multiplying by $-i$. 

The following are examples of eigenfunctions.
\begin{figure}[h!]
\center \includegraphics[width=6in]{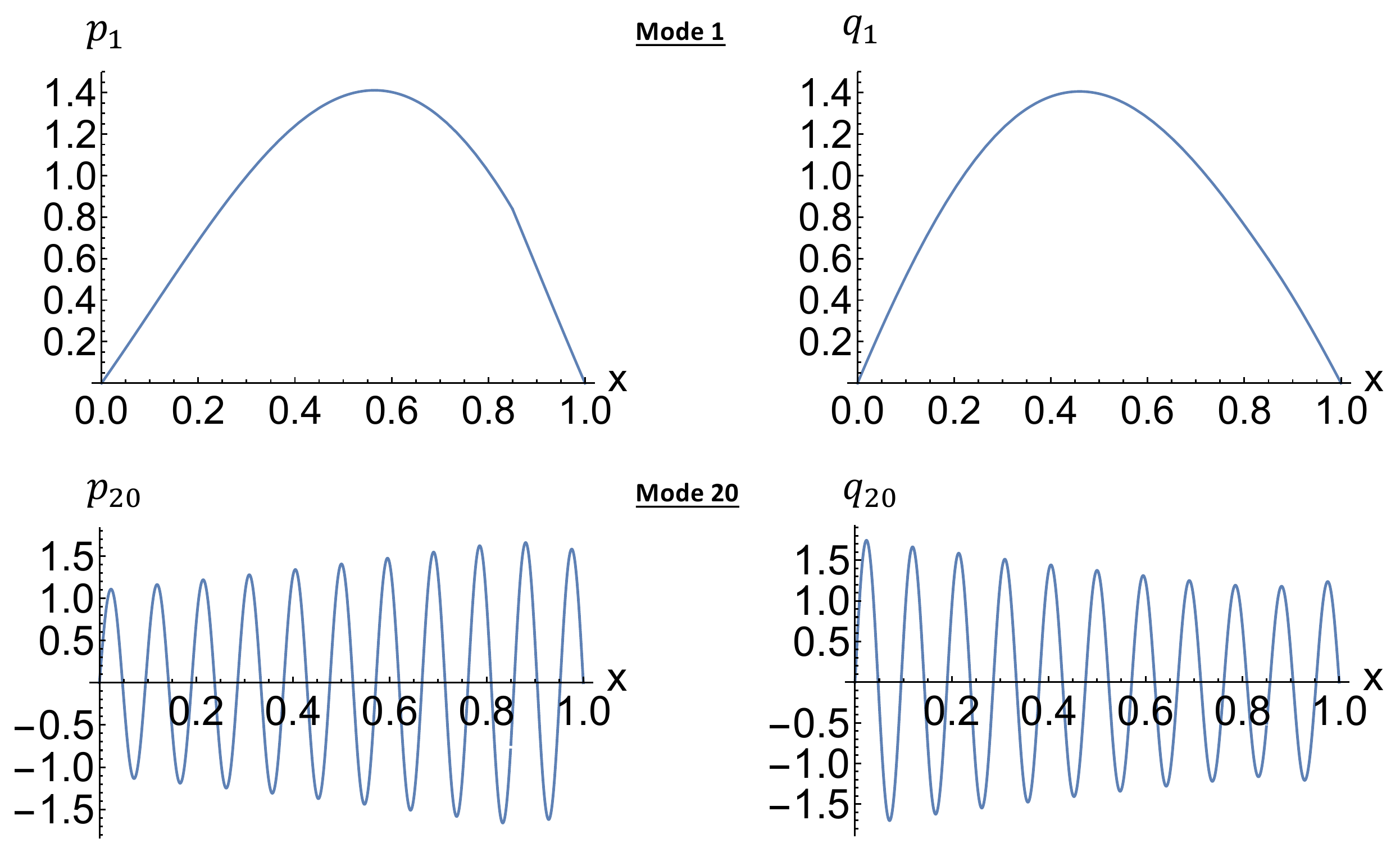}
\caption{The first and the twentieth modes for $X=0.85$, $v=1$. }
\end{figure}
\begin{figure}[h!]
\center \includegraphics[width=6in]{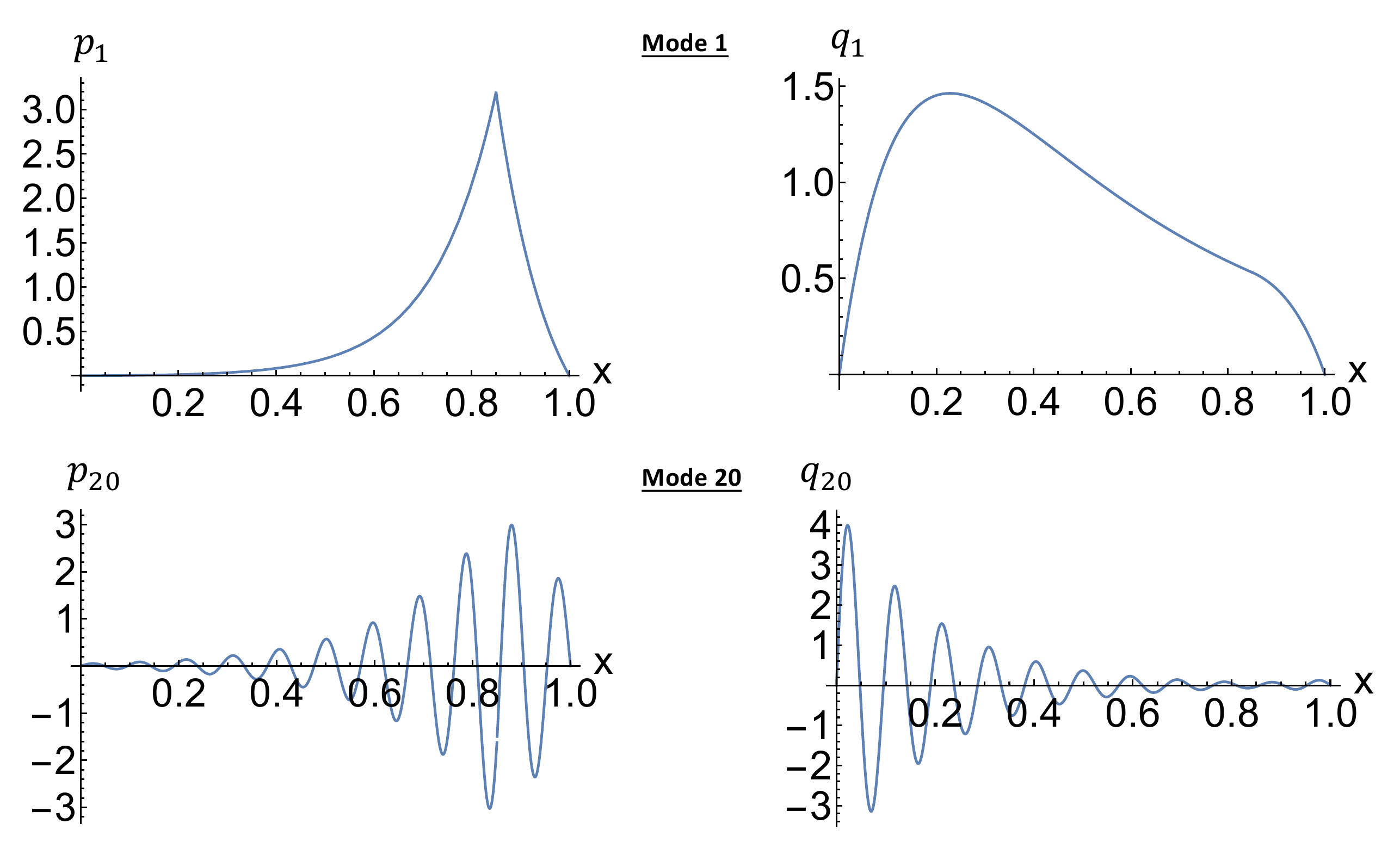}
\caption{The first and the twentieth modes for $X=0.85$, $v=10$. }
\end{figure}
The discontinuity in the slope of $p$s - but not of $q$s - is clearly visible in the first mode.  We can understand this by substituting the form $p(x) = q(x)e^{-U(x)}$ into the fourth boundary condition on $p$ (i.e. $vp_I(X) - \left.\frac{dp_I}{dx}\right|_{x=X} = -vp_{II}(X) - \left.\frac{dp_{II}}{dx}\right|_{x=X}$), and find that $\frac{dq}{dx}$ is continuous across the junction, i.e. $\left.\frac{q_I}{dx}\right|_{x=X} = \left.\frac{dq_{II}}{dx}\right|_{x=X}$.  The other three boundary conditions for $q$ are the same as for $p$.

With all this information, we conclude that the set of functions $\{q_n\}$ is then sufficient for us to be able to find the coefficients $c_n$ in the series  
$P(x,t) = \sum_{n} c_n p_n(x)e^{\sigma_n t}$ - as long as there is completeness.  The coefficients are given by 
\begin{equation}
c_n = \frac{\int_0^1 P(x,t=0) q^*_n(x) \,dx}{\int_0^1 q^*_n(x) p_n(x)\,dx}
\end{equation}
Completeness is not guaranteed, but unlike the two-layer case, we found that the method works, provided enough modes are used.  We will not discuss convergence properies of the series here.  

In relation to the mean first passage time problem, we are interested in the $\delta$-function initial condition, $P(x,t=0) = \delta(x-x_0)$, in which case the coefficients are given by 
\begin{equation}
c_n = \frac{q^*_n(x_0)}{\int_0^1 q^*_n(x) p_n(x)\,dx}.
\end{equation}

\newpage
\section*{C: Trajectory examples}
Fig.~4 and the subsequent discussion in our main text discussed several regimes of MFPT, depending on the value of $a=b$, for symmetric trap placement.  We now show trajectories in each of those regimes. 

First, we show trajectories in the plateau regime that precedes the second crossover.  This takes place for $a$ roughly in the range $[10^{-2},10]$.   This is shown in Fig.~\ref{fig:Trajectories_twolayer_lowest_a}.
\begin{figure}[h!]
\center \includegraphics[width=4in]{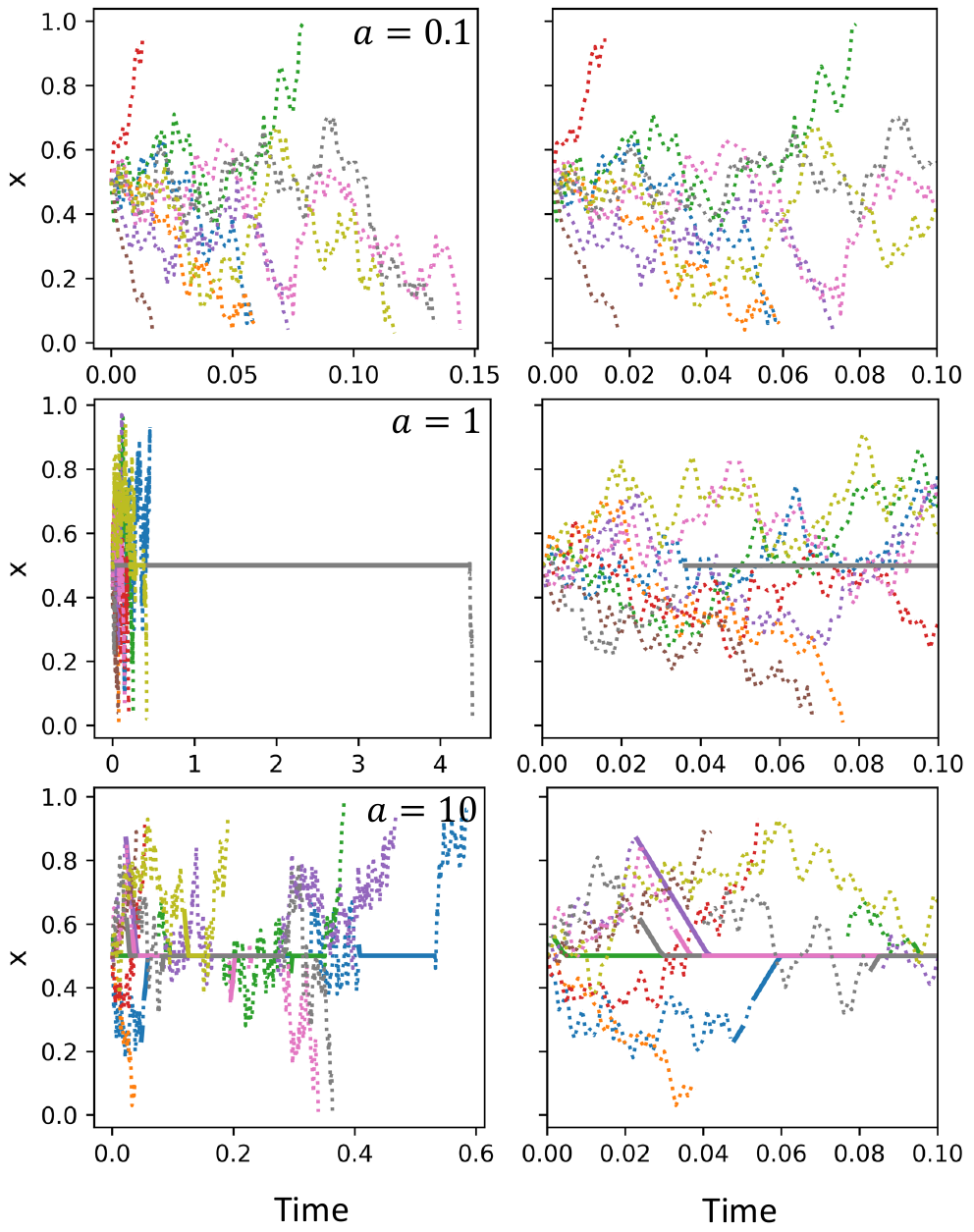}
\caption{Nine trajectories at lower $a$s.  All particles are placed initially at $x_0=0.5$ on the DL.  Here $X=0.5$ and $v=20$.  The right panels show a smaller window of time.}
\label{fig:Trajectories_twolayer_lowest_a}
\end{figure}

We can clearly see that as $a$ increases, thee probability of switching into the AL increases.  Once a particle switches to the AL, it will move towards the attractor.  

As $a$ increases further, the likelihood of the advective motion towards the attractor all in one ride on the AL decreases.  Instead, a typical particle will experience episodes of a little bit of advective motion, followed by a little bit of diffusive motion, and so on - see Fig.~\ref{fig:Trajectories_two_layer_middle_a}.  This happens in the second crossover regime that begins for $a\approx 10$ and continues for several decades.
\newpage
\begin{figure}[h!]
\center \includegraphics[width=4in]{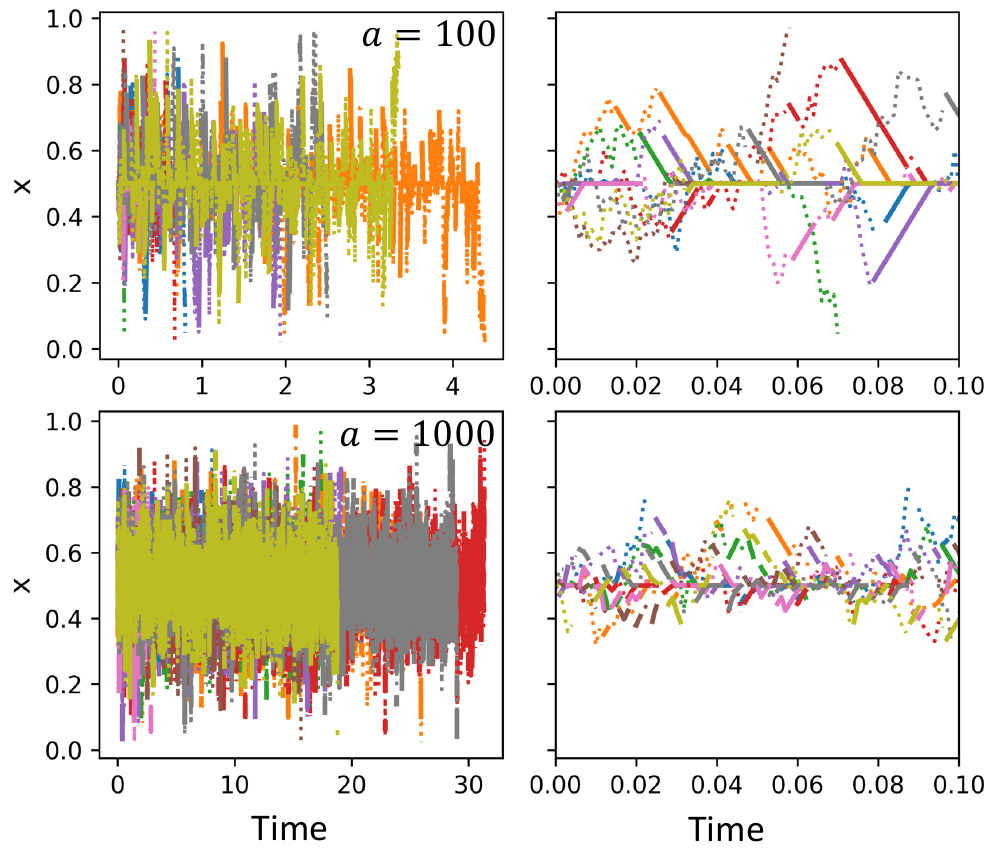}
\caption{Nine trajectories at intermediate $a$s.  All particles are placed initially at $x_0=0.5$.  Here $X=0.5$, $v=20$.  The right panels show a smaller window of time.}
\label{fig:Trajectories_two_layer_middle_a}
\end{figure}

For $a$ even larger - the system enters the second plateau, when any further increase in $a$ does not increase MFPT.  This means that the system behaves in accordance to the one-layer model \footnote{This is not what makes escape events rare.  The signature of the rarity of escape events  (that is MFPT is much greater than all other time scales) is the exponential growth of MFPT with $v$.}.  The the episodes of diffusion and advection become even shorter.  Trajectories in such a regime are shown in Fig.~\ref{fig:Trajectories_two_layer_higher_a}, for progressively narrower windows of time, from left to right.

\newpage
\begin{figure}[h!]
\center \includegraphics[width=6.5in]{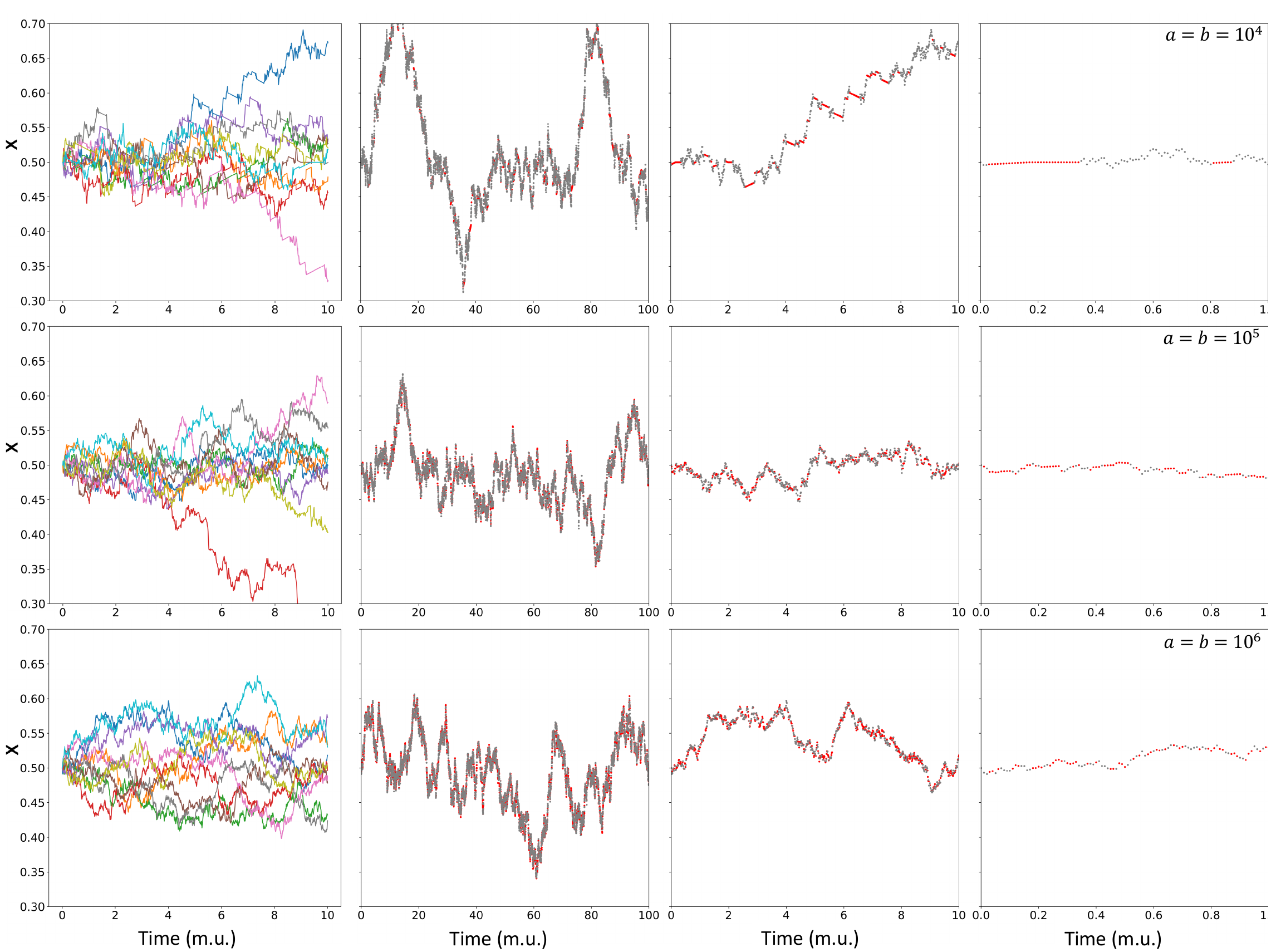}
\caption{Trajectories for $a$ between $10^4$ to $10^6$ in powers of $10$.  Here again $X=0.5$ and $v=20$.  Leftmost column has $10$ trajectories, while the other columns show one trajectory for progressively narrower windows of time, from left to right.  In these right three columns, the red color indicates advective portions of trajectories, while grey are diffusive portions. }
\label{fig:Trajectories_two_layer_higher_a}
\end{figure}

\newpage
\section*{D: Theory and simulation comparison - one-layer limit}
In this section we show the comparison between the one-layer analytical predictions of $p_l$, $p_r$, $\tau_l$, $\tau_r$, and $\tau$ with results of simulations of the two-layer model.
\begin{figure}[h!]
\center \includegraphics[width=5in]{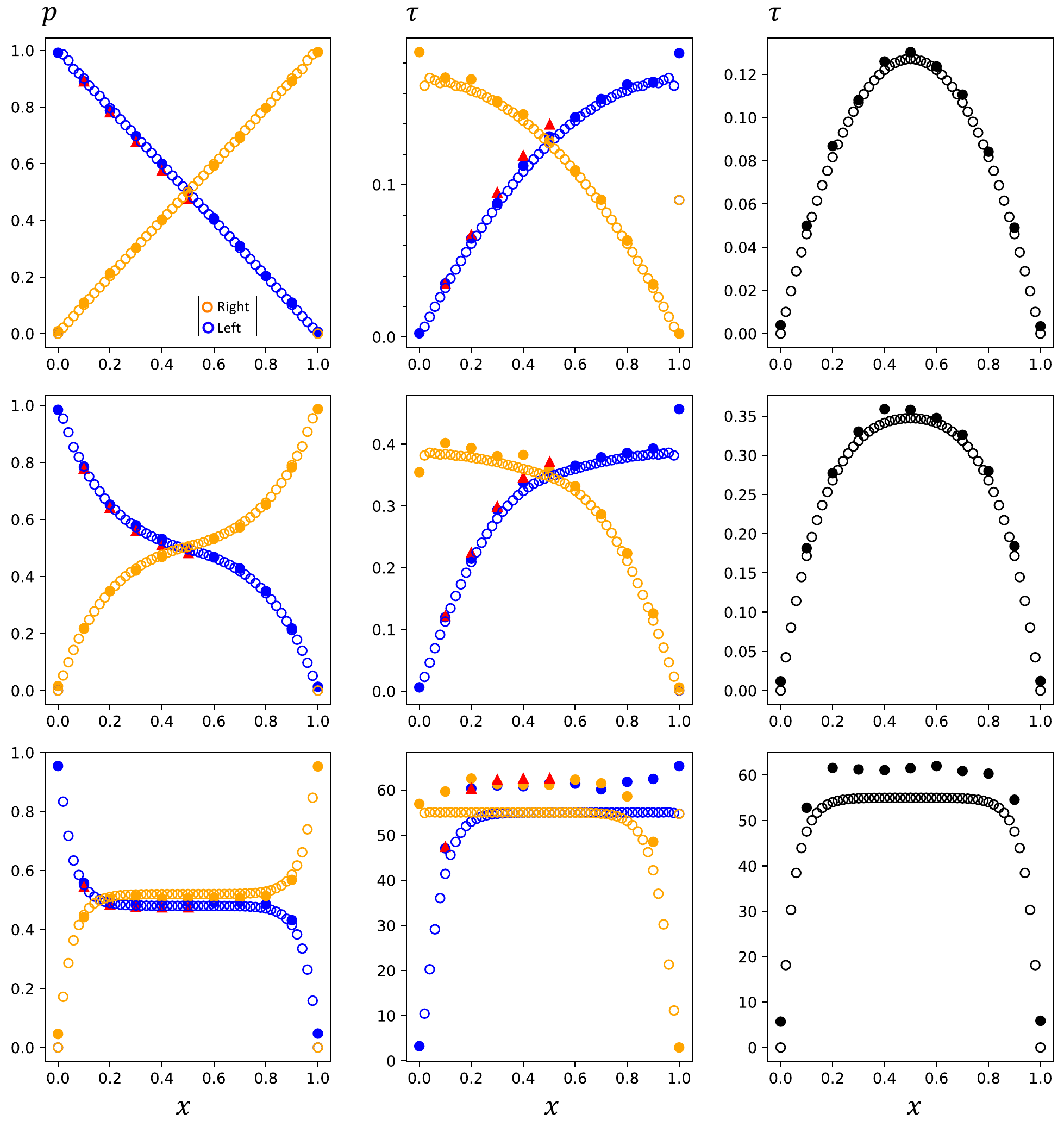}
\label{fig:matrix}
\caption{Comparison between analytical quantities (open circles) and simulation results (filled circles - Monte Carlo simulation as described in the main paper, filled triangles - forward flux sampling algorithm \cite{ffs}).  Left column: probabilities to escape through the left end (blue) $p_l$ and right end (orange) $p_r$.  Middle column: escape time conditioned on the left exit (blue) $\tau_l$ and right exit (orange) $\tau_r$.  Right column: net MFPT $\tau$.  The growing discrepancy between simulation and analytical results is due to the diffusive approximation of the latter; the details will be discussed in the coming publication \cite{Our_PRE}. Here $X=0.5$.  Top row: $v=0.1$, middle row $v=5$, bottom row $v=20$.}
\end{figure}
\newpage
\end{widetext}
\end{appendix}

\newpage
\bibliography{sample}

%

\end{document}